\documentclass[authoryear,12pt]{article}
\usepackage{amsmath}
\usepackage{times}
\usepackage{graphicx}
\usepackage{color}
\usepackage{multirow}
\usepackage{setspace}
\usepackage[authoryear]{natbib}
\usepackage{rotating}
\usepackage{bbm}
\usepackage{latexsym}
\usepackage[english]{babel}

\usepackage{subfig}
\usepackage{caption}
\usepackage{amssymb}
\usepackage{epsfig}

\begin{document}

\begin{titlepage}
\begin{center}
  {\LARGE Redundancy and synergy in dual decompositions of mutual information gain and information loss}
  \vspace*{5mm}

        {\large Daniel Chicharro $^{1,2,\ast}$ and Stefano Panzeri $^2$}

        \vspace*{7mm}
        {$^1$ \emph{Department of Neurobiology, Harvard Medical School, Boston, MA, USA}}\\[2mm]

        {$^2$ \emph{Neural Computation Laboratory, Center for Neuroscience and Cognitive Systems, Istituto Italiano di Tecnologia @UniTn, Rovereto (TN), Italy.}}\\[2mm]

        {$\ast$ Daniel\_Chicharro@hms.harvard.edu\ \ \ \ \ \ \ \ \ \ \ \ \ \ \ \ \ }\\[2mm]

\end{center}

\vspace*{3mm}

\begin{center}
{\large Abstract}
\end{center}

\cite{Williams10} proposed a nonnegative mutual information decomposition, based on the construction of information gain lattices, which allows separating the information that a set of variables contains about another into components interpretable as the unique information of one variable, or redundant and synergy components. In this work we extend the framework of \cite{Williams10} focusing on the lattices that underpin the decomposition. We generalize the type of constructible lattices and examine the relations between the terms in different lattices, for example relating bivariate and trivariate decompositions. We point out that, in information gain lattices, redundancy components are invariant across decompositions, but unique and synergy components are decomposition-dependent. Exploiting the connection between different lattices we propose a procedure to construct, in the general multivariate case, information decompositions from measures of synergy or unique information. We introduce an alternative type of mutual information decompositions based on information loss lattices, with the role and invariance properties of redundancy and synergy components exchanged with respect to gain lattices. We study the correspondence between information gain and information loss lattices and we define dual decompositions that allow overcoming the intrinsic asymmetry between invariant and decomposition-dependent components, which hinders the consistent joint characterization of synergy and redundancy.

\vspace*{3mm}


\end{titlepage}



\thispagestyle{empty}
%
\ \vspace{-0mm}\\

\pagenumbering{arabic}
\section{Introduction}
\label{s1}

The aim to determine the mechanisms producing dependencies in a multivariate system, and to characterize these dependencies, has motivated several proposals to breakdown the contributions to the mutual information between sets of variables \citep{Timme14}. This problem is interesting from a theoretical perspective in information theory, but it is also crucial from an empirical point of view in many fields of systems and computational biology \citep[e.\ g.\ ][]{Anastassiou07, Ludtke08, Watkinson09, Oizumi14, Faes16}. For example, in neuroscience breaking down the contributions to mutual information between sets of variables is fundamental to make any kind of progress in understanding neural population coding of sensory information. This breakdown is in fact necessary to identify the unique contributions of individual classes of neurons, and of interactions among them, to the sensory information carried by neural populations \citep{ Averbeck06, Panzeri15}, is necessary to understand how information in populations of neurons contributes to behavioural decisions \citep{Haefner13, Panzeri17}, and to understand how information is transmitted and further processed across areas \citep{Wibral14}.

Consider the mutual information $I(\mathbf{S};\mathbf{R})$ between two possibly multivariate sets of variables $\mathbf{S}$ and $\mathbf{R}$, here thought, for the sake of example, as a set of sensory stimuli, $\mathbf{S}$, and neural responses $\mathbf{R}$, but generally any sets of variables. An aspect that has been widely studied is how dependencies within each set contribute to the information. For example, the mutual information breakdown of \cite{Panzeri99, Pola03} quantifies the global contribution to the information of conditional dependencies between the variables in $\mathbf{R}$, and has been applied to study how interactions among neurons shape population coding of sensory information. Subsequent decompositions, based on a maximum entropy approach, have proposed to subdivide this contribution separating the influence of dependencies of different orders \citep{Amari01, Ince10}. However, these types of decompositions do not ensure that all terms in the decomposition are nonnegative and hence should be better interpreted as a comparison of the mutual information across different alternative system's configurations \citep{Latham05, Chicharro14b}. Two concepts tightly related to this type of decompositions are those of redundancy and synergy \citep[e.\ g.\ ][]{Schneidman03}. Redundancy refers to the existence of common information about $\mathbf{S}$ that could be retrieved from different variables contained in $\mathbf{R}$ used separately. Conversely, synergy refers to the existence of information that can only be retrieved when jointly using the variables in $\mathbf{R}$. Traditionally, synergy and redundancy had been quantified together, with the measure called interaction information \citep{McGill54} or co-information \citep{Bell03}. A positive value of this measure is considered as a signature of redundancy being present in the system, while a negative value is associated with synergy, so that redundancy and synergy have traditionally been considered as mutually exclusive.

The seminal work of \cite{Williams10} introduced a new approach to decompose the mutual information into a set of nonnegative contributions. Let us consider first the bivariate case. Without loss of generality, from now on we assume $S$ to be a univariate variable, if not stated otherwise. For the bivariate case \cite{Williams10} argued that the mutual information can be decomposed into four terms:

\begin{equation}
I(S;12) =I(S;1.2)+I(S;1 \backslash 2)+I(S;2 \backslash 1)+I(S;12 \backslash 1,2).
\label{i000}
\end{equation}
The term $I(S;1.2)$ refers to a redundancy component between variables $1$ and $2$. The terms $I(S;1 \backslash 2)$ and $I(S;2 \backslash 1)$ quantify a component of the information that is unique of $1$ and of $2$, respectively, that is, some information that can be obtained from one of the variables alone but that cannot be obtained from the other alone. The term $I(S;12 \backslash 1,2)$ refers to the synergy between the two variables, the information that is unique for the joint source $12$ with respect to the variables alone. Note that in this decomposition a redundancy and a synergy component can exist simultaneously. In fact, \cite{Williams10} showed that the measure of co-information is equivalent to the difference between the redundancy and the synergy terms of Eq.\ \ref{i000}. Generally, \cite{Williams10} defined this type of decomposition for any multivariate set of variables $\{ \mathbf{R} \}$. The key ingredients for this general formulation were the definition of a general measure of redundancy and the association of each decomposition comprising $n$ variables to a lattice structure, constructed with different combinations of groups of variables ordered by defining an ordering relation. We will review this general formulation linking decompositions and lattices in great detail below.

Different parts of the framework introduced by \cite{Williams10} have generated different levels of consensus. The conceptual framework of nonnegative decompositions of mutual information, with distinguishable redundancy and synergy contributions and with lattices underpinning the decompositions, has been widely accepted. Conversely, it has been argued that the specific measure $I_{min}$ originally used to determine the terms of the decomposition does not properly quantify redundancy \cite[e.\ g.\ ][]{ Harder12, Griffith13}. Accordingly, much of the subsequent efforts have focused in finding the right measures to define the components of the decomposition. From these alternative proposals, some take as the basic component to derive the terms in the decomposition another measure of redundancy \citep{Harder12, Ince16}, but also a measure of synergy \citep{Griffith13}, or of unique information \citep{Bertschinger12}. In contrast to $I_{min}$, these measures fulfill the identity axiom \citep{Harder12}, introduced to prevent that for $\mathbf{S}$ composed by two independent variables, a redundancy component is obtained for $\mathbf{R}$ being a copy of $\mathbf{S}$. Indeed, apart from proposing other specific measures, subsequent studies have proposed a set of axioms which state desirable properties of these measures \citep{Griffith13, Harder12, Rauh14, Griffith14}. However, there is no full consensus on which are the axioms that should be imposed. Furthermore, it has been shown that some of these axioms are incompatible with each other \citep{Rauh14}. In particular, \cite{Rauh14} provided a counterexample illustrating that nonnegativity is not ensured for the components of the decomposition in the multivariate case if assuming the identity axiom. Some contributions have also studied the relation between the measures that contain different number of variables \citep{Bertschinger12b, Rauh14}. For some specific type of variables, multivariate Gaussians with a univariate $S$, the equivalence between some of the proposed measure has been proven \citep{Barret15}.

To our knowledge, perhaps because of these difficulties in founding a proper measure to construct the decompositions, less attention has been paid to study the properties of the lattices associated with the decompositions. We here focus on examining these properties and the basic constituents that are used to construct the decompositions from the lattices. We generalize the type of lattices introduced by \cite{Williams10} and we examine the relation between the information-theoretic quantities associated with different lattices (Section \ref{s2_1}). Since one of the challenges when using other proposed measures to construct the decompositions has been the extension to the multivariate case \citep[e.\ g.\ ][]{Griffith13, Bertschinger12}, we consider how to identify the terms in the decomposition when using as a basic component a measure of synergy or unique information (Section \ref{s2_2}). Motivated by this analysis, we introduce a new type of lattices, namely information loss lattices in contrast to the information gain lattices described in \cite{Williams10}. We show that these loss lattices are more naturally related to synergy measures, as opposed to gain lattices more naturally related to redundancy measures (Section \ref{s3}). Finally, we identify the existence of dual information gain and loss lattices, which share the basic terms of the decompositions and have desirable consistency properties that allow properly characterizing redundancy and synergy simultaneously (Section \ref{s4}). Other open questions related to the selection of the measures and the axioms are out of the scope of this work.

We now continue with the revision of the decompositions of \cite{Williams10} as a first step for the extensions we propose in this work. For the bivariate decomposition of Eq.\ \ref{i000}, the terms in the decomposition of $I(S;12)$ can be readily related to the marginal and conditional mutual informations consistently. In particular, given the usual information-theoretic relations \citep{Cover06}

\begin{subequations}
\begin{align}
I(S;12) &= I(S;1)+I(S;2|1)\\
&= I(S;2)+I(S;1|2)
\end{align}
\label{i00}
\end{subequations}
we see that

\begin{equation}
I(S;1) = I(S;1.2)+I(S;1 \backslash 2)
\label{i01}
\end{equation}
and

\begin{equation}
I(S;2|1) = I(S;2 \backslash 1)+I(S;12 \backslash 1,2),
\label{i02}
\end{equation}
and analogously for $I(S;2)$ and $I(S;1|2)$. That is, each variable can contain some information that is redundant to the other and some part that is unique. Conditioning one variable on the other removes the redundant component of the information but adds the synergistic component, resulting in the conditional information being the sum of the unique and synergistic terms.

We now review the construction of the lattices and their relation to the decompositions. A lattice is composed by a set of \emph{collections}. This set is defined as

\begin{equation}
\mathcal{A}(\mathbf{R})= \{ \alpha \in \mathcal{P}(\mathbf{R}) \backslash \{ \emptyset \}: \forall\ \mathbf{A}_i, \mathbf{A}_j \in \alpha, \mathbf{A}_i \nsubseteq \mathbf{A}_j\},
\label{i1}
\end{equation}
where $\mathcal{P}(\mathbf{R}) \backslash \{ \emptyset \}$ is the set of all nonempty subsets of the set of nonempty \emph{sources} that can be formed from $\{ \mathbf{R} \}$, where a source $\mathbf{A}$ is a subset of the variables $\{ \mathbf{R} \}$. That is, each collection $\alpha$ is itself a set of sources, and each source $\mathbf{A}$ is a set of variables. The domain of the collections included in the lattice is established by the constraint that a collection cannot contain sources that are a superset of another source in the collection. This restriction is justified in detail in \cite{Williams10}, based on the idea that the redundancy between a source and any superset of it is equal to the information of that source. Given the set of collections $\mathcal{A}(\mathbf{R})$, the lattice is constructed defining an ordering relation between the collections, which becomes meaningful for the decomposition because redundancy monotonically increases in agreement with the ordering relation \citep[see Theorem 2 in][]{Williams10}. In particular:

\begin{figure}
  \begin{center}
    \scalebox{0.5}{\includegraphics*{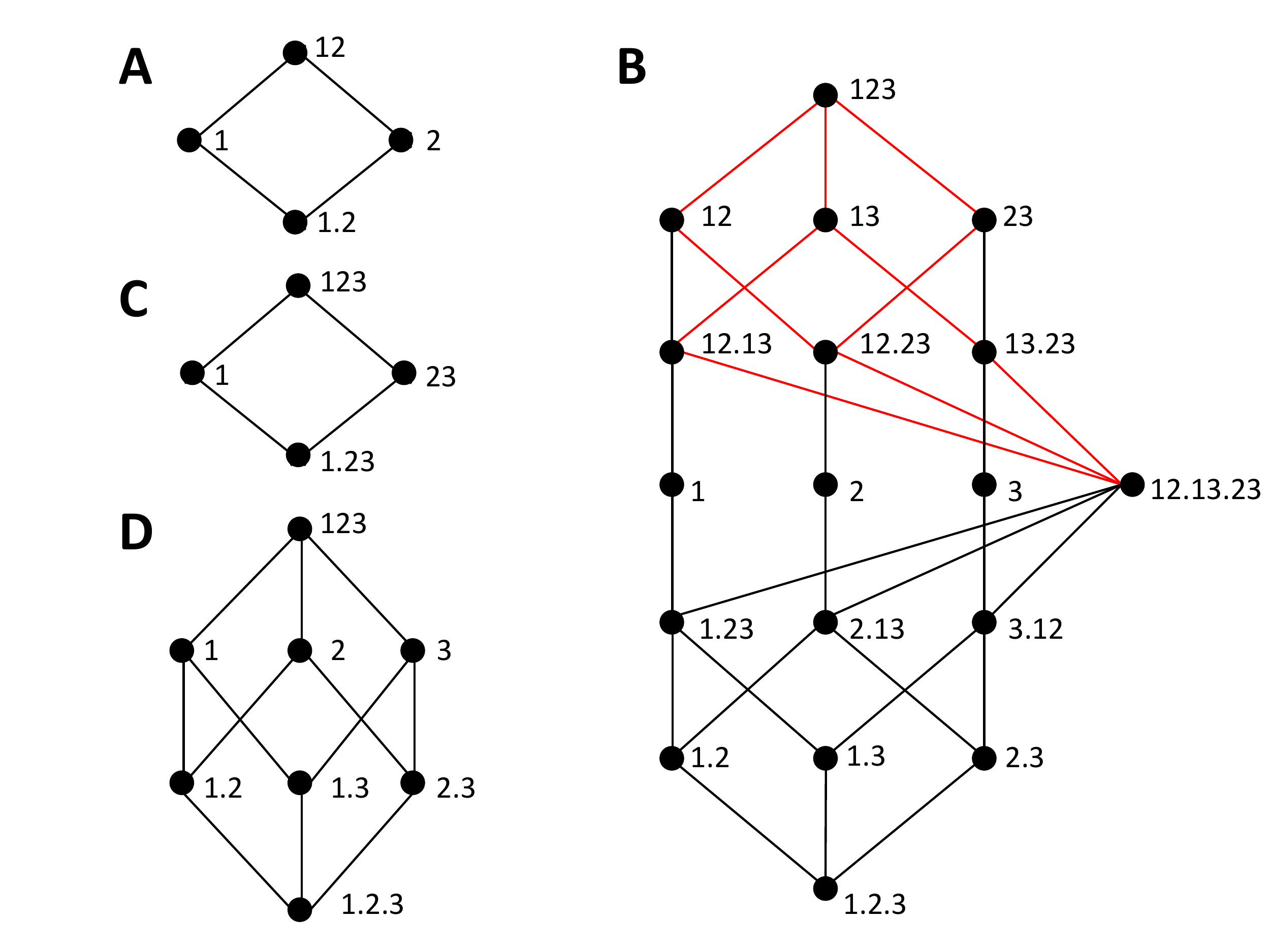}}
  \end{center}
  \caption{Information gain decompositions of different orders and for different subsets of collections of sources. \textbf{A}), \textbf{B}) Lattices constructed from the complete domain of collections as defined by Eq.\ \ref{i1} for $n=2$ and $n=3$, respectively. Red edges in \textbf{B}) identify the embedded lattice formed by collections that do not contain univariate sources. \textbf{C}) Alternative decomposition based only on sources $1$ and $23$. \textbf{D}) Alternative decomposition that does not contain bivariate sources.}
  \label{fig1}
\end{figure}

\begin{equation}
\forall\ \alpha, \beta \in \mathcal{A}(\mathbf{R}), (\alpha \preceq \beta \Leftrightarrow \forall \mathbf{B} \in \beta, \exists \mathbf{A} \in \alpha, \mathbf{A} \subseteq \mathbf{B}),
\label{i2}
\end{equation}
that is, for two collections $\alpha$ and $\beta$, $\alpha \preceq \beta$ if for each source in $\beta$ there is a source in $\alpha$ that is a subset of that source. The lattices constructed for the case of $n=2$ and $n=3$ using this ordering relation are shown in Figure \ref{fig1}A,B. In this work we use a different notation than in \cite{Williams10}, which allows us to shorten a bit the expressions. For example, instead of writing $\{1\}\{23\}$ for the collection composed by the source containing variable $1$ and the source containing variables $2$ and $3$, we write $1.23$, that is, we save the curly brackets that indicate for each source the set of variables and we use instead a dot to separate the sources.

Each collection in the lattice is associated with a measure of the redundancy between the sources composing the collection. \cite{Williams10} defined a measure of redundancy, called $I_{min}$, that is well defined for any collection. In this work we do not need to consider the specific definition of $I_{min}$. What is relevant for us is that, when ascending the lattice, $I_{min}$ monotonically increases, being a cumulative measure of information and reaching the total amount of information at the top of the lattice. Based on this accumulation of information, we will from now on refer to the type of lattices introduced by \cite{Williams10} as \emph{information gain lattices}. Furthermore, we will generically refer to the terms quantifying the information accumulated in each collection as \emph{cumulative terms} and denote the cumulative term of a collection $\alpha$ by $I(S,\alpha)$. The reason for this change of terminology will become evident when we introduce the information loss lattices, since redundancy is not specific of the information gain lattices, and thus it is more appropriate to disentangle it nominally from the cumulative terms, even if in the formulation of \cite{Williams10} they are inherently associated.

The mutual information decomposition was constructed in \cite{Williams10} by implicitly defining partial information measures associated with each node, such that the cumulative terms are obtained from the sum of partial information measures:

\begin{equation}
I(S,\alpha) = \sum_{\beta \in \downarrow \alpha} \Delta_{\mathcal{C}}(S;\beta).
\label{i3}
\end{equation}
In particular, $\downarrow\alpha$ refers to the set of collections lower than or equal to $\alpha$, given the ordering relation (see Appendix $A$ for details). Again, here we will adopt a different terminology and we will refer to $\Delta_{\mathcal{C}}(S;\beta)$ as the \emph{incremental term} of the collection $\beta$ in lattice $\mathcal{C}$, instead of as the partial information measure. This is because, as we will see, it is convenient to consider incremental terms as increments that can equally be of information gain or information loss. As proved in \cite[][Theorem $3$]{Williams10}, Eq.\ \ref{i3} can be inverted to:

\begin{subequations}
\begin{align}
\Delta_{\mathcal{C}}(S;\alpha) &= I(S;\alpha)- \sum_{k=1}^{|\alpha^-|} (-1)^{k-1} \sum_{
\begin{array}{c} \mathcal{B} \subseteq \alpha^- \\ |\mathcal{B}|=k \end{array}} \sum_{\beta \in \bigcap\limits_{\gamma \in \mathcal{B}}\downarrow \gamma} \Delta_{\mathcal{C}}(S;\beta) \\
&= I(S,\alpha)- \sum_{k=1}^{|\alpha^-|} (-1)^{k-1} \sum_{
\begin{array}{c} \mathcal{B} \subseteq \alpha^- \\ |\mathcal{B}|=k \end{array}} I(S; \bigwedge \mathcal{B}),
\end{align}
\label{i4}
\end{subequations}
where $\alpha^-$ is the cover set of $\alpha$ and $\bigwedge \mathcal{B}$ is the infimum of the set $\mathcal{B}$ (see Appendix $A$ for details).

\section{Extended information gain decompositions from redundancy, uniqueness or synergy measures}
\label{s2}

In this section we still focus on the information gain decompositions introduced by \cite{Williams10}. In Section \ref{s2_1} we motivate the extension of their approach to comprise a more general set of lattices, built based on subsets of the domain of collections determined in Eq.\ \ref{i1}. We examine the validity of each lattice construction depending on the properties and relations of the variables involved and we study the relation between the terms of different lattices. In Section \ref{s2_2} we address how to calculate the terms of multivariate decompositions associated with information gain lattices when the basic measure that is defined \emph{a priori} is a measure of synergy or unique information, instead of a measure of redundancy (which directly identifies the cumulative terms of the lattice). Our analysis indicates some inconsistencies in the simultaneous characterization of synergy and redundancy in multivariate systems and leads to the introduction of information loss lattices in Section \ref{s3} and ultimately to the characterization of dual information gain and information loss lattices in Section \ref{s4}.

\subsection{Relations between information gain decompositions with different subsets of sources collections}
\label{s2_1}

\cite{Williams10} studied how to decompose the mutual information in decompositions composed by all the collections of sources in $\mathcal{A}(\mathbf{R})$. Figure \ref{fig1}A-B show the corresponding lattices for $n=2$ and $n=3$, respectively. However, the number of collections in these decompositions rapidly increases when the number of variables increases (e.g. $7579$ collections for $n=5$), which may render the decompositions difficult to handle in practice. Here we generalize their approach in a straightforward way, considering decompositions composed by any subset $\mathcal{C} \subseteq \mathcal{A}(\mathbf{R})$ which elements still form a lattice (see Appendix $C$ for a discussion of more general decompositions based on subsets that do not form a lattice). For example, Figure \ref{fig1}C shows the decomposition formed by the collections that combine the sources $1$ and $23$. In Figure \ref{fig1}B the red edges indicate the decomposition based on collections combining the sources $12$, $13$, $23$, without further decomposing the contribution of single variables separately. Oppositely, Figure \ref{fig1}D shows the decomposition based on the sources $1$, $2$, and $3$, which does not include bivariate sources resulting from merging these univariate sources. A certain decomposition can be embedded within a bigger one, as indicated in Figure \ref{fig1}B, but generally considering more collections alters the structure of the lattice, by modifying the cover relations between the nodes. For example, the decomposition of Figure \ref{fig1}D is not embedded in Figure \ref{fig1}B. Similarly, the cover relations in the bivariate decomposition of $\mathcal{A}(\{1,2\})$ in Figure \ref{fig1}A change in the trivariate decomposition of $\mathcal{A}(\{1,2,3\})$ in Figure \ref{fig1}B since nodes $12.13$ and $12.23$ appear between $12$ and $1$ and $2$, respectively. The same occurs between $1.2$ and $1$ and $2$, with nodes $1.23$ and $2.13$, respectively. Furthermore, the down set of $12$, in comparison to the bivariate lattice, comprises others nodes because of the presence of $12.13.23$.

When studying multivariate systems, the nature and relation between the variables may provide some \emph{a priori} information in favor of a certain decomposition. For example, in the case of Figure \ref{fig1}C, variables $2$ and $3$ can correspond to two signals recorded from the same subsystem, while $1$ is a signal from a different subsystem. This may render a bivariate decomposition more adequate, even if having three variables. For example, this is a common scenario when recording brain signals from different brain areas and the analysis of interactions can be carried out at different spatial scales \citep{Panzeri15}. Similarly, in the case of Figure \ref{fig1}D, one may prefer to simplify the analysis without explicitly considering all synergistic contributions of bivariate sources. Another possibility is that, even if it is known that a system is composed by a certain number of variables, only a subset is available for the analysis, and it is thus important to understand how the influence of the missing variables is reflected in each term of the decomposition (e.g. how the terms in the full decomposition for $n=3$ that contain $1$ and $2$ are merged in the fewer terms of the full decomposition of $n=2$). Again this is a common scenario when studying neural population coding of sensory stimuli, since usually only simultaneous recordings from a subset of the neural population, or from one of the brain regions involved, is available. In any case, in order to better choose the most useful decomposition given a certain set of concrete variables, and to understand how the different decompositions are related, we need to consider how the terms from one decomposition are mapped to another.

The connection between the terms in two different decompositions is qualitatively different for the cumulative terms, $I(S,\alpha)$, and the incremental terms $\Delta_{\mathcal{C}}(S;\alpha)$. A cumulative term $I(S,\alpha)$ quantifies the information about $S$ that is redundant within a certain collection of sources $\alpha$. This information is well defined without considering which is the set $\mathcal{C}$ of collections that has been selected, that is, it depends only on $S$ and $\alpha$. Accordingly, the cumulative terms of information gain $I(S,\alpha)$ are invariant across decompositions. Oppositely, as we here explicitly indicate in our notation, the incremental terms $\Delta_{\mathcal{C}}(S;\alpha)$ are in general decomposition-dependent. This can be seen from Eq.\ \ref{i4}: the cumulative terms used to calculate $\Delta_{\mathcal{C}}(S;\alpha)$ depend on the specific structure of the lattice, in particular on which is the increment sublattice $\diamondsuit \alpha$ in that lattice (See Appendix $A$ for details). This is summarized indicating that:

\begin{equation}
I(S;\alpha) = I_{\mathcal{C}}(S;\alpha)= I_{\mathcal{C}'}(S;\alpha), \forall \mathcal{C}, \mathcal{C}',
\label{s2e1}
\end{equation}
while for the incremental terms only a sufficient condition for equality across decompositions can be formulated:

\begin{equation}
\diamondsuit_{\mathcal{C}} \alpha = \diamondsuit_{\mathcal{C}'} \alpha \Rightarrow \Delta_{\mathcal{C}}(S;\alpha) = \Delta_{\mathcal{C}'}(S;\alpha),
\label{s2e2}
\end{equation}
which is a direct consequence of Eq.\ \ref{s2e1} given the dependence of the incremental terms on the cumulative terms (Eq.\ \ref{i4}).

\begin{figure}
  \begin{center}
    \scalebox{0.5}{\includegraphics*{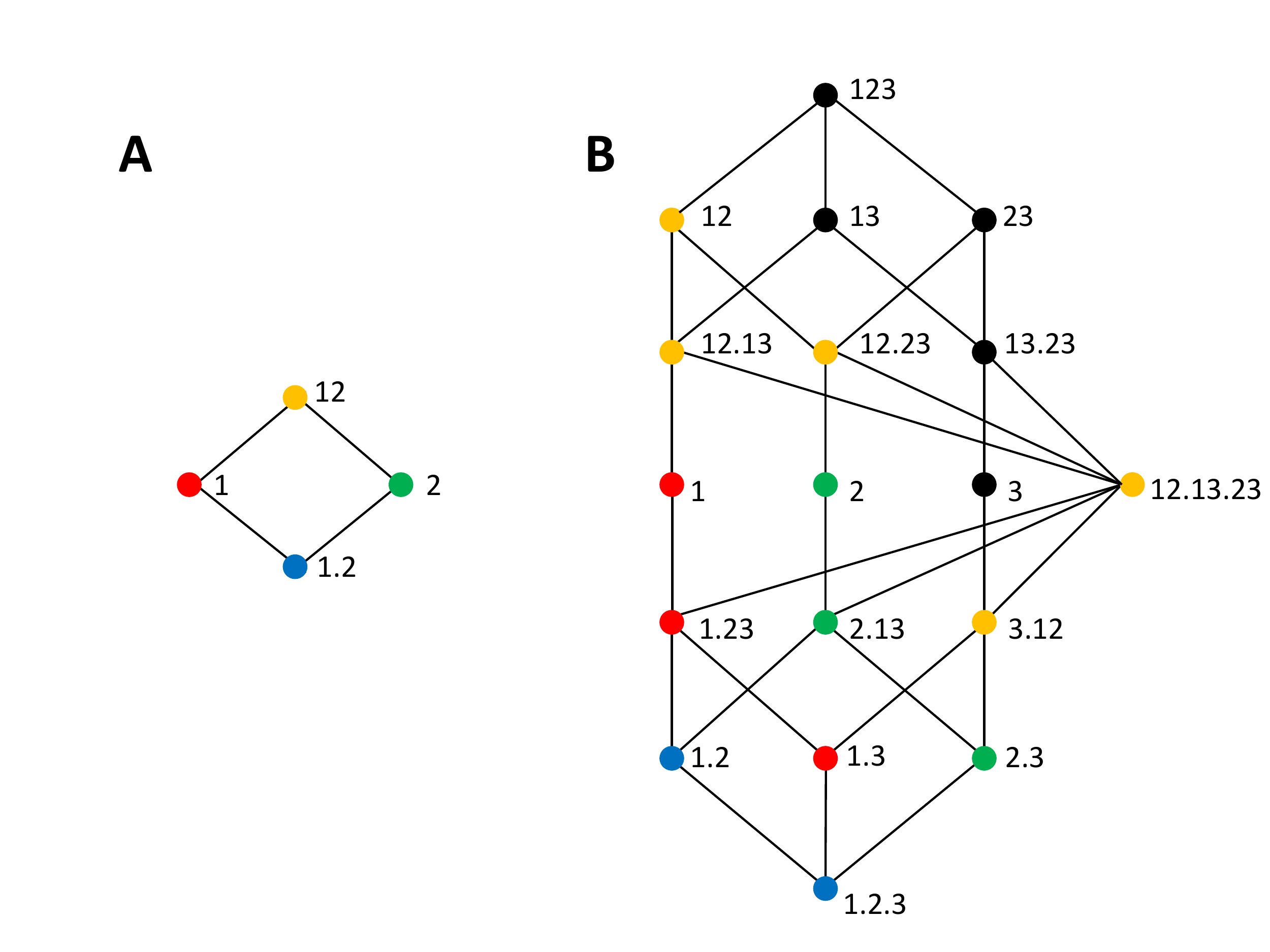}}
  \end{center}
  \caption{Mapping of the incremental terms of the bivariate lattice for $1$, $2$ to the full trivariate lattice for $1$, $2$, $3$. \textbf{A}) The bivariate lattice with each incremental term marked with a different color. \textbf{B}) The trivariate lattice with the incremental terms in which each of the incremental terms of the bivariate lattice is subdecomposed indicated with the same color.}
  \label{fig1B}
\end{figure}

Each cumulative term that is present in two decompositions provides an equation that relates the incremental terms in those decompositions, since in each lattice cumulative terms result from the accumulation of increments according to Eq.\ \ref{i3}. In particular, for two decompositions $\mathcal{C}$ and $\mathcal{C}'$ with a common collection $\alpha$

\begin{equation}
 I(S;\alpha) = \sum_{\beta \in \downarrow_{\mathcal{C}} \alpha} \Delta_{\mathcal{C}}(S;\beta) = \sum_{\beta \in \downarrow_{\mathcal{C}'} \alpha} \Delta_{\mathcal{C}'}(S;\beta).
\label{s2e3}
\end{equation}
In general, these type of relations impose some constraints that involve several incremental terms from each decomposition. In the cases in which a decomposition is composed by a set of collections $\mathcal{C}'$ which is a subset of another set $\mathcal{C}$, then combining these constraints allows decomposing each of the incremental terms of the subsumed set $\mathcal{C}'$ as a sum of incremental terms of the bigger set $\mathcal{C}$. For example, when connecting the incremental terms of the decompositions of Figure \ref{fig1}A,C, we get only the constraint $\Delta_A(S;1)+\Delta_A(S;1.2)=\Delta_B(S;1)+\Delta_B(S;1.23)$, given the only common node $I(S;1)$. Conversely, the set $\mathcal{A}(\{1,2\})$ of the decomposition of Figure \ref{fig1}A is a subset of $\mathcal{A}(\{1,2,3\})$ in Figure \ref{fig1}B, and the constraints allow detailing each incremental term of the decomposition with $\mathcal{A}(\{1,2\})$ as the sum of several terms of the decomposition with $\mathcal{A}(\{1,2,3\})$, as shown in Figure \ref{fig1B}.

As a last general point regarding the possibility to choose different decompositions when a set of variables is available, we indicate that, when deterministic relationships exist between the variables, the definition of the domain of collections (Eq.\ \ref{i1}) and the ordering relation (Eq. \ref{i2}) can impose some limitations on the decompositions that are possible. In particular, Eq.\ \ref{i1} excludes any collection in which a source is a superset of any other. Consider for example the case of three variables $1,2,3$ such that $12$ completely determine $3$. Accordingly, in the decomposition of Figure \ref{fig1}B several collections are altered, since the source $12$ could be replaced by $123$. This leads to the presence of invalid collections in the set, such as $123.13$ instead of $12.13$, since $13$ is a subset of $123$. Similarly, given the deterministic relation, one could reduce $123$ to $12$, duplicating this last collection and affecting the ordering relation of the top element with $13$ and $23$. In general, this means that a certain lattice cannot be taken as valid a priori. Conversely, it should be verified, for each specific set of variables, if the collections that compose it are valid once the properties of the variables are taken into account.

The exclusion of certain lattices in the presence of deterministic relations can be seen as a limitation of the decomposition framework, but on the other hand this verification turns out to be important to avoid problematic cases. In particular, it allows avoiding the counterexample provided in \cite{Rauh14} to show that it is not always possible when $n>2$, independently of how the terms of the decomposition are defined, to obtain a nonnegative decomposition. This counterexample is based on three variables such that any pair deterministically determines the third. Without using any specific definition of the measures associated with the nodes the authors proved that at least a certain incremental term of the lattice of Figure \ref{fig1}B is negative in this case. However, given the deterministic relations between the variables, all the collections comprising a bivariate source need to be excluded from the set, since these bivariate sources are equivalent to the source $123$ and thus any other source in the collection is a subset of this one. Similarly, $123$ can be reduced to any collection containing a single source composed by a pair of the variables, which duplicates these collections and affects the ordering relations. In Appendix $B$ we show in more detail that when reconsidering the counterexample of \cite{Rauh14} for decompositions that comply the constraints of Eqs. \ref{i1} and \ref{i2} the existence of a negative term does not hold anymore. Therefore, our extended approach, which generally considers alternative decompositions compatible with a set of variables, can overcome the limitations of adopting the unique lattice $\mathcal{A}(\mathbf{R})$ for each set of variables $\mathbf{R}$ with $n=|\mathbf{R}|$. However, note that the possibility to deal with cases like the one raised in \cite{Rauh14} by adapting the lattice does not preclude from the potential existence of negative incremental terms. As we reviewed above, the definition of the proper measure for the decompositions is an open question, and finding a measure that ensures generally the nonnegativity of the incremental terms, or identifying the properties of the variables or the decompositions that ensures this nonnegativity, is out of the scope of this work. Only in Appendix $C$ we review the requirements to obtain nonnegative incremental terms. For this purpose we reexamine from a more general perspective several of the Theorems of \cite{Williams10}, identifying the key ingredients of the proofs that are sustained by lattice properties, general properties captured in the axioms that have been proposed \citep{Harder12}, or by properties specific of their measure $I_{min}$.

\subsection{The determination of information gain cumulative terms from synergy or unique information measures by combining information gain decompositions of different orders}
\label{s2_2}

Above we have examined possible alternative decompositions of mutual information gain and the relations among them from a generic perspective, based only on the structure of the decompositions and the general properties of the cumulative and incremental terms. Now we discuss more specifically how the expressions corresponding to each term can be found given a specific measure that is defined as the basis to construct the decomposition. If a measure of accumulated mutual information gain $I(S,\alpha)$ is defined, it is straightforward to calculate all the terms in the decomposition. This is the case of the seminal work of \cite{Williams10}, where the cumulative terms were defined as the redundancy measures $I_{min}$. Once the cumulative terms have been calculated, the incremental terms can be calculated using Eq.\ \ref{i4}.

However, the calculation of all terms is not so straightforward if the measure defined as the basis to construct the decomposition does not define the cumulative terms. In fact, in the different proposals that exist so far, the basic component chosen to calculate the other terms has alternatively been a redundancy measure \citep{Williams10, Harder12, Ince16}, a synergy measure \citep{Griffith13}, or a unique information measure \citep{Bertschinger12}. In the bivariate case, these alternatives do not lead to any qualitative difference in the procedure to identify the other measures, because, given Eqs.\ \ref{i000} and \ref{i01}, we can relate the four terms of the bivariate decomposition with $I(S;1)$, $I(S;2)$, and $I(S;12)$, so that defining one of the four terms is enough to identify the other three. However, this direct procedure cannot similarly be applied for $n>2$. This can be understood, already for $n=3$, considering the number of cumulative terms which are directly calculable as mutual information terms. Only the terms related to the collections formed by a single source, $1$, $2$, $3$, $12$, $13$, $23$, and $123$, are defined \emph{a priori}. This means that only seven equations analogous to Eqs.\ \ref{i000} and \ref{i01} are available to calculate the $K=18$ cumulative terms. If the measure taken as basis of the decomposition is defined generally for each node (as $I_{min}$ in \cite{Williams10}) this is not a problem, and these seven equations are directly fulfilled as special cases of Eq.\ \ref{i3}. But if the measure taken as the basis is a measure of synergy or uniqueness, then it does not define directly the cumulative terms, but only certain incremental terms. This difference is clear already for $n=2$. The redundancy $I(S;1.2)$ is a cumulative term in the decomposition, in particular it corresponds to the bottom element of the lattice. Conversely, the unique information terms $I(S; 1 \backslash 2)$ and $I(S; 2 \backslash 1)$, as well as the synergy $I(S; 12 \backslash 1,2)$ correspond to incremental terms. That is, the particularity of the redundancy measure $I_{min}$ is that it provides a definition for all the cumulative terms of the mutual information gain decomposition, while the measures of unique information or synergy, for $n>2$, do not provide a definition applicable to all the incremental terms of the lattice. Indeed, previous approaches based on synergy or unique information measures have not provided a general procedure to determine the expression of all the elements in multivariate decompositions.

We will now indicate how to calculate all the cumulative terms of the mutual information gain decomposition for $n=3$ using as a basis a definition of synergy or unique information. As we will show below, this procedure can lead to some inconsistencies, but it serves to motivate the introduction of the alternative decompositions of the mutual information loss instead of the mutual information gain. The key ingredient here is the invariance of the cumulative terms across decompositions, as indicated in Eq.\ \ref{s2e1}. Based on this invariance we can resort to the bivariate decompositions in order to calculate many of the cumulative terms of the trivariate decomposition of Figure \ref{fig1}B. Indeed, from the $18$ minus $7$ terms that do not correspond directly to the mutual information of a single source, all except the ones of the collections $12.13.23$ and $1.2.3$ appear also in a bivariate decomposition. For example, $1.2$ is part of the decomposition in Figure \ref{fig1}A, and $1.23$ is part of the one in Figure \ref{fig1}C. Analogous bivariate decompositions exist for $1.3$, $2.3$, $2.13$, $3.12$, $12.13$, $12.23$, and $13.23$. For each of these bivariate decompositions, if a definition of bivariate synergy is defined, it can be used to determine the corresponding bivariate redundancy, which, being a cumulative term, is invariant and can be used equally in the trivariate decomposition. Accordingly, it is the connection between different decompositions what allows us to calculate most of the terms. This same procedure of using the bivariate decompositions could be used if instead of a definition of synergy we used a definition of unique information. Finally, to calculate $1.2.3$ and $12.13.23$ we can use the smaller trivariate decompositions of Figure \ref{fig1}D and the one composed by the red edges of Figure \ref{fig1}B, respectively. In these two smaller trivariate decompositions, after using the bivariate ones to calculate the corresponding cumulative terms, the situation becomes the same as for the bivariate case: all cumulative terms are already calculated except one, which means that it suffices to define a single measure, either a synergy or unique information, in order to be able to retrieve the complete set of cumulative and incremental terms.

\begin{figure}
  \begin{center}
    \scalebox{0.5}{\includegraphics*{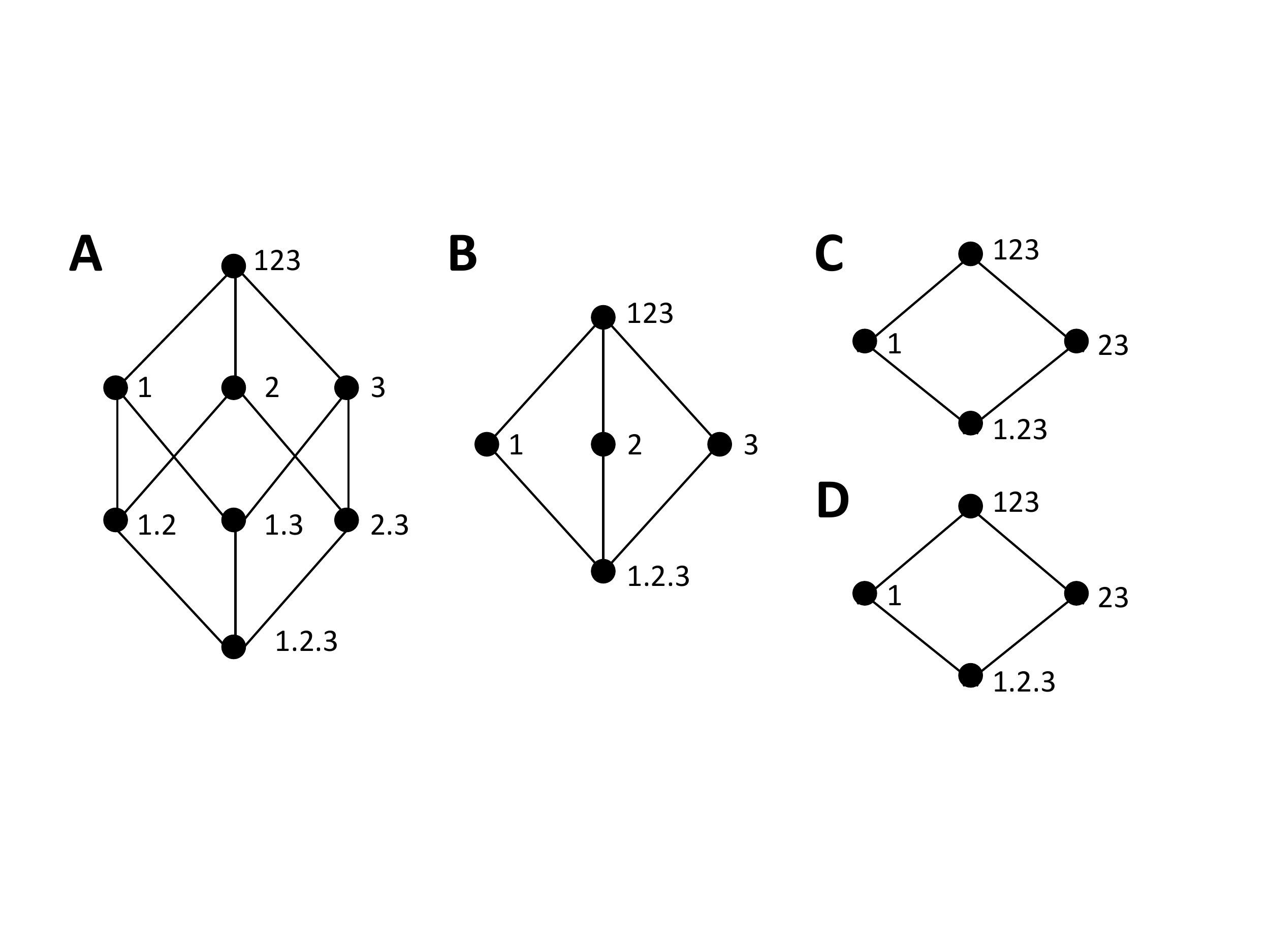}}
  \end{center}
  \caption{Examples of information gain lattices that result in inconsistencies when trying to derive redundancy terms from a synergy definition, as explained in Section \ref{s2_2}.}
  \label{fig2}
\end{figure}

This procedure is attractive because, nicely using only the connection between different lattices and the invariance of the cumulative terms, it apparently provides a way to construct multivariate decompositions, simply by recurrently using decompositions of a lower order to calculate the cumulative terms. However, this approach leads to some inconsistencies. In particular, consider that a measure of synergy is provided, which should allow identifying the top incremental term of any decomposition. For example, a measure of synergy should determine the incremental term $\Delta(S; 123 \backslash 12.13.23)$ of Figure \ref{fig1}B and $\Delta(S; 123 \backslash 1.2.3)$ of Figure \ref{fig1}D. However, since in a decomposition of the mutual information gain these synergy components correspond to incremental terms, as discussed above, they are decomposition-specific. Consider then the alternative decompositions presented in Figure \ref{fig2}A-B, and Figure \ref{fig2}C-D, respectively. In Figure \ref{fig2}A-B, if the same definition of $\Delta(S; 123 \backslash 1.2.3)$ is used based on the synergy measure, this results in a different form for $I(S;1.2.3)$ in the two lattices, because the increment sublattices $\diamondsuit_{A} 123 \neq \diamondsuit_{B} 123$. This contradicts the invariance of the cumulative terms across lattices. Figure \ref{fig2}C-D presents another contradiction resulting from directly using the synergy definition: since collections $123$, $1$ and $23$ are common to both decompositions, the same expression would be obtained for the redundancy $I(S; 1.23)$ and $I(S; 1.2.3)$ depending on the lattice used. For both examples the problem is that a definition of synergy is expected to depend only on $S$ and on the sources among which synergy is quantified, but cannot be context-dependent, in opposition to the incremental terms, which are always context-dependent in the sense that they are decomposition-specific.

\section{Decompositions of mutual information loss}
\label{s3}

A further problem raised by the comparisons in Figure \ref{fig2} is the following: if conversely to having a definition of synergy, a measure defining the cumulative terms is used as in the original proposal of \cite{Williams10}, the incremental terms are calculated using Eq.\ \ref{i4}, and when the increment sublattices of the top incremental term are different, different quantifications associated with synergy are obtained. That is, for example, $\Delta_A(S; 123 \backslash 1.2.3) \neq \Delta_B(S; 123 \backslash 1.2.3)$ in Figure \ref{fig2}. Accordingly, it is not straightforward to interpret the top incremental term as the one quantifying the synergistic component of the mutual information, since different possible decompositions result in different terms. This issue does not arise for the bivariate decomposition because a single decomposition involving a synergistic component is possible.

\begin{figure}
  \begin{center}
    \scalebox{0.55}{\includegraphics*{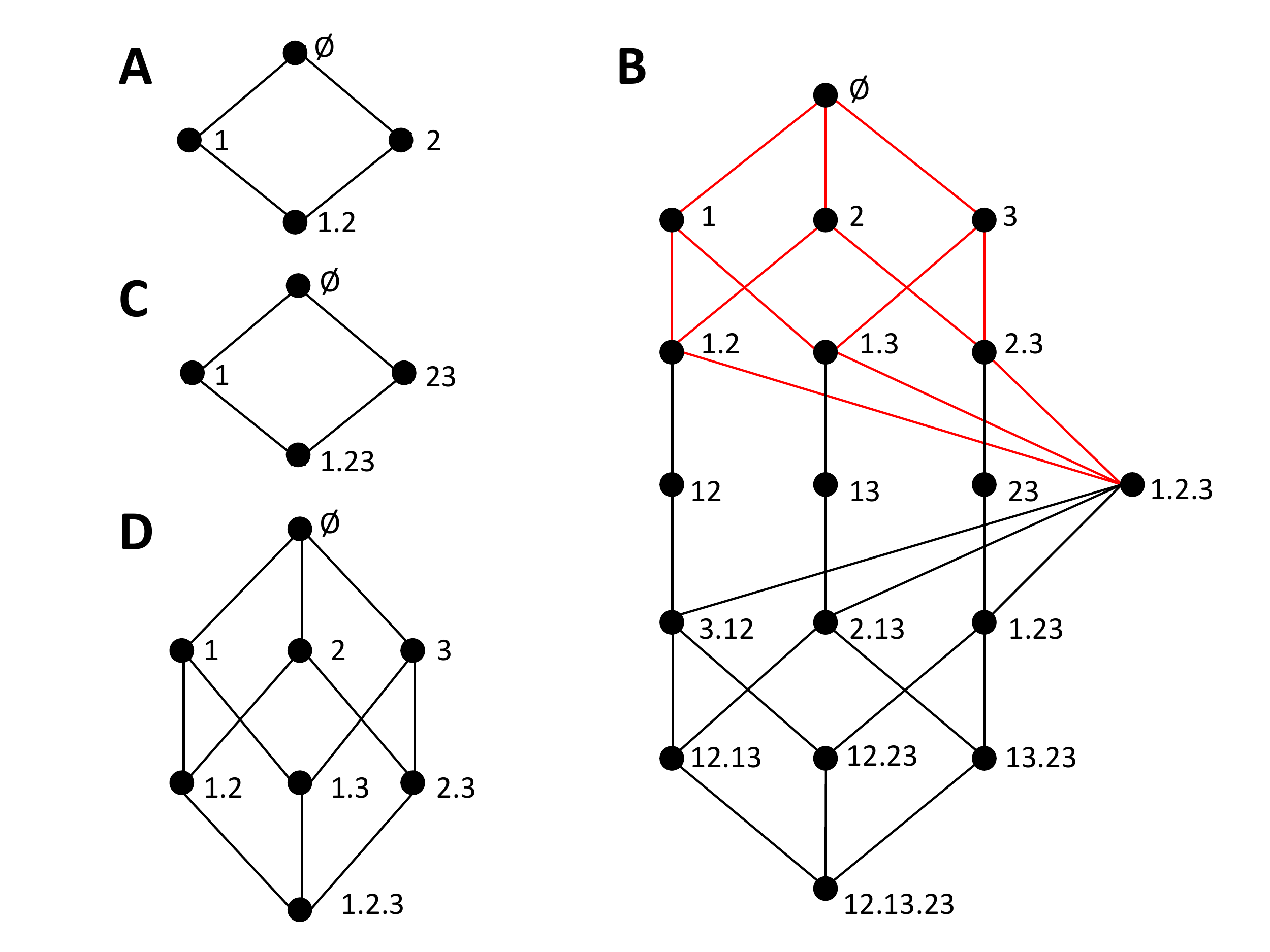}}
  \end{center}
  \caption{Information loss decompositions of different orders and for different subsets of collections of sources. The lattices are analogous to the information gain lattices of Figure \ref{fig1}. Note that now the lattice embedded in \textbf{B}) which is indicated with the red edges corresponds to the one shown in \textbf{D}).}
  \label{fig3}
\end{figure}

To overcome these problems, we now consider an alternative type of decompositions: decompositions of mutual information loss instead of decompositions of mutual information gain. In this type of decompositions, synergy measures can be associated with cumulative terms instead of incremental terms, and thus they are not decomposition-specific. In the lattices associated with the decompositions of mutual information gain, the ordering relation is defined such that upper nodes correspond to collections which cumulative terms have more information about $S$ than each of the cumulative terms in their down set. Oppositely, in the lattice associated with a decomposition of mutual information loss, an upper node corresponds to a higher loss of the total information contained in the whole set of variables about $S$. The domain of the collections valid for the information loss decomposition can be defined analogously to the case of information gain:

\begin{equation}
\mathcal{A}^*(\mathbf{R})= \{ \alpha \in \mathcal{P}(\mathbf{R}) \backslash \{ \mathbf{R} \}: \forall\ \mathbf{A}_i, \mathbf{A}_j \in \alpha, \mathbf{A}_i \nsubseteq \mathbf{A}_j\}.
\label{s3e1}
\end{equation}
Note that this domain is equivalent to the one of the information gain decompositions (Eq.\ \ref{i1}), except that the collection corresponding to the source containing all variables $\{ \mathbf{R} \}$ is excluded instead of the empty collection. This because, in the same way that no information gain can be accumulated with no variables, no loss can be accumulated with all variables. Furthermore, $\mathcal{A}^*(\mathbf{R})$ excludes collections that contain sources that are supersets of other sources of the collection, equally to $\mathcal{A}(\mathbf{R})$. An ordering relation is also introduced analogously to Eq.\ \ref{i2}:

\begin{equation}
\forall\ \alpha, \beta \in \mathcal{A}^*(\mathbf{R}), (\alpha \preceq \beta \Leftrightarrow \forall \mathbf{B} \in \beta, \exists \mathbf{A} \in \alpha, \mathbf{B} \subseteq \mathbf{A}).
\label{s3e2}
\end{equation}
This ordering relation differs from the one of lattices associated with information gain decompositions in that now upper collections should contain subset sources and not the opposite. Figure \ref{fig3} shows several information loss decompositions analogous to the gain decompositions of Figure \ref{fig1}. For the lattices of Figure \ref{fig3}A,C,D, the only difference with respect to Figure \ref{fig1} is the top node, where the collection containing all variables is replaced by the empty set. Indeed, the empty set results in the highest information loss. For the full trivariate decomposition of Figure \ref{fig3}B there are many more changes in the structure of the lattice with respect to Figure \ref{fig1}B. In particular, now the smaller embedded lattice indicated with the red edges corresponds to the one of Figure \ref{fig3}D, while the lattice of Figure \ref{fig1}D is not embedded in Figure \ref{fig1}B. An intuitive way to interpret the mutual information loss decomposition is in terms of the marginal probability distributions from which information can be obtained for each collection of sources. Each source in a collection indicates a certain probability distribution that is available. For example, the collection $12.13$, composed by the sources $12$ and $13$, is associated with the preservation of the information contained in the marginal distributions $p(S, 1, 2)$ and $p(S, 1, 3)$. Note that all distributions are joint distributions of the sources and $S$. In this view, the extra information contained in $p(S; \mathbf{R})$ that cannot be obtained from the marginals preserved, corresponds the accumulated information loss. Accordingly, the information loss decompositions can be connected to hierarchical decompositions of the mutual information \citep{Olbrich15, Perrone16}. Furthermore, information loss associated with the preservation of only certain marginal distributions can be formulated in terms of maximum entropy \citep{Bertschinger12}, which renders loss lattices suitable to extend previous work studying neural population coding with the maximum entropy framework \citep{Ince10}.

We will use the notation $L(S;\alpha)$ to refer to the cumulative terms of the information loss decomposition, in comparison to the cumulative terms of information gain $I(S;\alpha)$. For the incremental terms, since they also correspond to a difference of information (in this case lost information) we will use the same notation. This will be further justified below when examining the dual relationship between certain information gain and loss lattices. However, when we want to explicitly indicate the type of lattice to which an incremental term belongs, we will explicitly distinguish $\Delta I$ and $\Delta L$. Importantly, the role of synergy measures and redundancy measures is exchanged in the information loss lattice with respect to the information gain lattice. In particular, in the information loss lattices the bottom element of the lattice corresponds to the synergistic term that in the information gain lattices is located at the top element. This represents a qualitative difference because now it is the synergy measure which is associated with cumulative terms, and redundancy is quantified by an incremental term. For example, in Figure \ref{fig3}B, $L(S;12.13.23)$ quantifies the information loss of considering only the sources $12.13.23$ instead of the joint source $123$, which is a synergistic component. On the other hand, the incremental term $\Delta(S; \emptyset \backslash 1,2,3)$ quantifies the information loss of either removing the source $1$ or, removing $2$, or removing $3$. Since the information loss quantified is associated with removing any of these sources, it means that the loss corresponds to information which was redundant to these three sources. This reasoning applies also to identify the uniqueness nature of other incremental terms of the information loss lattice. For example, $\Delta (S; 12.13 \backslash 23)$ can readily be interpreted as the unique information contained in $23$ that is lost when having only sources $12.13$.

The definition of the information loss lattices simplifies the construction of mutual information decompositions from a synergy measure. If such a measure can generically be used to define the cumulative terms of the loss lattice analogously to how a redundancy measure, for example $I_{min}$, defines the cumulative terms of the gain lattice, then the equations relating cumulative and incremental terms can be applied to identify all the remaining terms. Since, like in the information gain lattice, the top cumulative term is equal to the total information ($L(S; \emptyset)= I(S; \{\mathbf{R}\})$), the lattice is a decomposition of the mutual information for a certain set of variables $\{\mathbf{R}\}$. In particular, the relations between cumulative and incremental terms are totally equivalent to the ones of the information gain lattices:

\begin{equation}
L(S,\alpha) = \sum_{\beta \in \downarrow \alpha} \Delta_{\mathcal{C}}L(S;\beta),
\label{s3e3}
\end{equation}
and

\begin{subequations}
\begin{align}
\Delta_{\mathcal{C}}L(S;\alpha) &= L(S;\alpha)- \sum_{k=1}^{|\alpha^-|} (-1)^{k-1} \sum_{
\begin{array}{c} \mathcal{B} \subseteq \alpha^- \\ |\mathcal{B}|=k \end{array}} \sum_{\beta \in \bigcap\limits_{\gamma \in \mathcal{B}}\downarrow \gamma} \Delta_{\mathcal{C}}L(S;\beta) \\
&= L(S,\alpha)- \sum_{k=1}^{|\alpha^-|} (-1)^{k-1} \sum_{
\begin{array}{c} \mathcal{B} \subseteq \alpha^- \\ |\mathcal{B}|=k \end{array}} L(S; \bigwedge \mathcal{B}).
\end{align}
\label{s3e4}
\end{subequations}
The introduction of information loss lattices solves the problem of the ambiguity of the synergy terms derived from information gain lattices, which was caused by the identification of synergistic contributions with incremental terms, which are decomposition-specific by construction. In the information loss decomposition the synergy contributions are identified with cumulative terms, and thus are not decomposition-specific. Note however, that there is still a difference between the degree of invariance of the cumulative terms in the information gain decompositions and in the information loss decompositions. The loss is \emph{per se} relative to a maximum amount of information that can be achieved. This means that the cumulative terms of the information loss decomposition are only invariant across decompositions that have in common the set of variables from which the collections are constructed. This asymmetry between the absolute and relative nature of information gain and information loss is reflected in the following relations, which indicate how a single node $\alpha$ partitions between gain and loss the total information in each of the two types of lattices:

\begin{subequations}
\label{s3e5}
\begin{align}
I(S;\alpha) &= \sum_{\beta \in \downarrow \alpha} \Delta I(S;\beta) \\
I(S;\{ \mathbf{R}\})-I(S;\alpha) &= \sum_{\beta \in (\downarrow \alpha)^C} \Delta I(S;\beta)\\
L(S;\alpha') &= \sum_{\beta \in \downarrow \alpha'} \Delta L(S;\beta)\\
I(S;\{ \mathbf{R}\})-L(S;\alpha') &= \sum_{\beta \in (\downarrow \alpha')^C} \Delta L(S;\beta),
\end{align}
\end{subequations}
where $(\downarrow \alpha)^C = \mathcal{C} \backslash \downarrow \alpha$ is the complementary set to the down set of $\alpha$ given the particular set of collections $\mathcal{C}$ used to build a lattice. These equations indicate that in the information gain lattice all nodes (collections) out of the down set of $\alpha$ correspond to the information not gained by $\alpha$, or equivalently, to the information loss by using $\alpha$ instead of the whole set of variables. Analogously, in the information loss lattice, all nodes out of the down set of $\alpha'$ contain the information not lost by $\alpha'$, i.e., the information gained by using $\alpha'$. Accordingly, in both types of lattices we can say that each collection $\alpha$ partitions the lattice into an accumulation of gained and lost information.

\begin{figure}
  \begin{center}
    \scalebox{0.5}{\includegraphics*{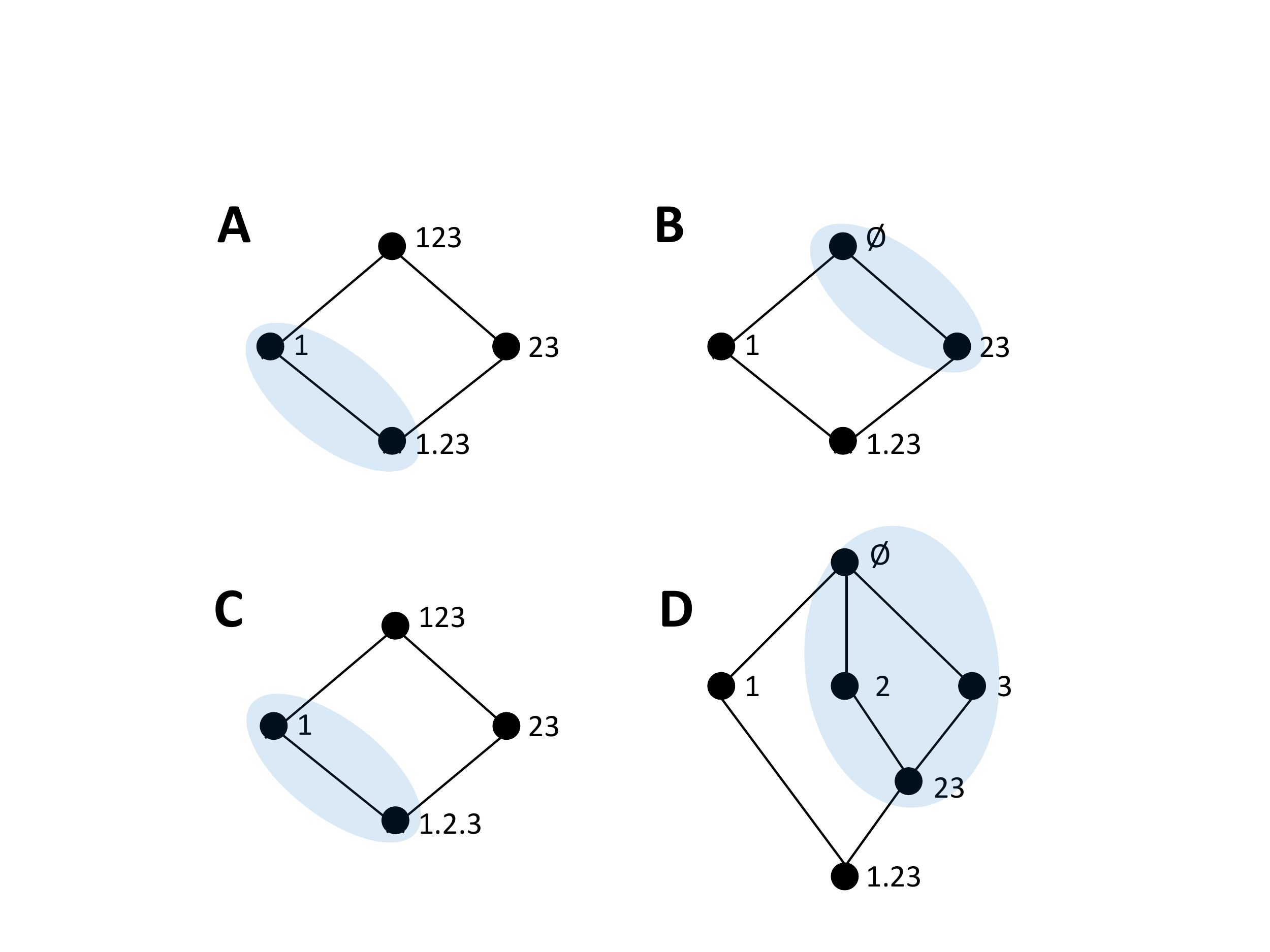}}
  \end{center}
  \caption{The correspondence between information gain and information loss lattices. \textbf{A}, \textbf{C}) Examples of information gain lattices, and their paired information loss lattices (\textbf{B}, \textbf{D}, respectively). The shaded areas comprise the collections corresponding to incremental terms that contribute to $I(S;1)$ in each lattice.}
  \label{fig4}
\end{figure}

\section{Dual decompositions of information gain and information loss}
\label{s4}

Comparing the information gain lattices and the information loss lattices we see that the former seem adequate to quantify unambiguously redundancy and the latter to quantify unambiguously synergy. In the same way that in relation to Figure \ref{fig2} we discussed that the top incremental terms of different lattices can have different values and do not correspond to a unique quantification of synergistic contributions, equivalently for the information loss lattices the top incremental elements generally differ across lattices, and cannot be associated with a unique quantification of redundancy contributions. Therefore, trying to retrieve the terms of information loss lattices from a definition of a redundancy measure, using a procedure analogous to the one discussed in Section \ref{s2_2}, would lead to the same kind of inconsistencies. We would like to understand in more detail, given Eqs.\ \ref{s3e5}, how the two types of lattices are connected, i.e., which relations exist between the cumulative or incremental terms of each other, and how to quantify synergy and redundancy together.


To address these questions we start indicating that, while in some cases it seems possible to establish a connection between the components of a pair composed by an information gain and an information loss lattice, in other cases the lack of a match is immediately evident. Consider the examples of Figure \ref{fig4}. In Figure \ref{fig4}A,C we reconsider the information gain lattices of Figure \ref{fig2}C,D, which we examined in Section \ref{s2_2} to illustrate that we arrive to an inconsistency when trying to extract the bottom cumulative term from the directly calculable mutual informations of $1$ and $23$ and a definition of synergy. Figure \ref{fig2}B,D show information loss lattices candidates to be associated with these gain lattices, respectively, based on the correspondence of the bottom and top collections. While the two information gain lattices only differ from each other in the bottom collection, the information loss lattices are substantially more different, with a different number of nodes. This occurs because, as we discussed above, the concept of a redundancy $1.2.3$ is associated with a loss that is common to removing any of the three variables, considered as the only source of information, and thus a separation of $2$ and $3$ from the source $23$ is required to quantify this redundancy.

The fact that the information gain lattice of Figure \ref{fig4}C and the information loss lattice of Figure \ref{fig4}D have a different number of nodes already indicates that a complete match between their components is not possible. For example, consider the decomposition of $I(S;1)$ in the information gain lattice, as indicated by the nodes comprised in the shaded area in Figure \ref{fig4}C. $I(S;1)$ is decomposed into two incremental terms. To understand which nodes are associated with $I(S;1)$ in the information loss lattice we argue, based on Eq.\ \ref{s3e5}d, that since the node $1$ is related to the accumulated loss $L(S;1)= I(S;123)-I(S;1)$, and $L(S;\emptyset)= I(S;123)$, this means that the sum of all the incremental terms which are not in the down set of $1$ must correspond to $I(S;1)$. These nodes are indicated by the shaded area in Figure \ref{fig4}D. Clearly, there is no match between the incremental terms of the information gain lattice and of the information loss lattice, since in the former $I(S;1)$ is decomposed into two incremental terms and in the latter is decomposed into four incremental terms. Conversely, for the lattices of Figure \ref{fig4}A,B, the number of incremental terms is the same, which does not preclude from a match.

\begin{figure}
  \begin{center}
    \scalebox{0.5}{\includegraphics*{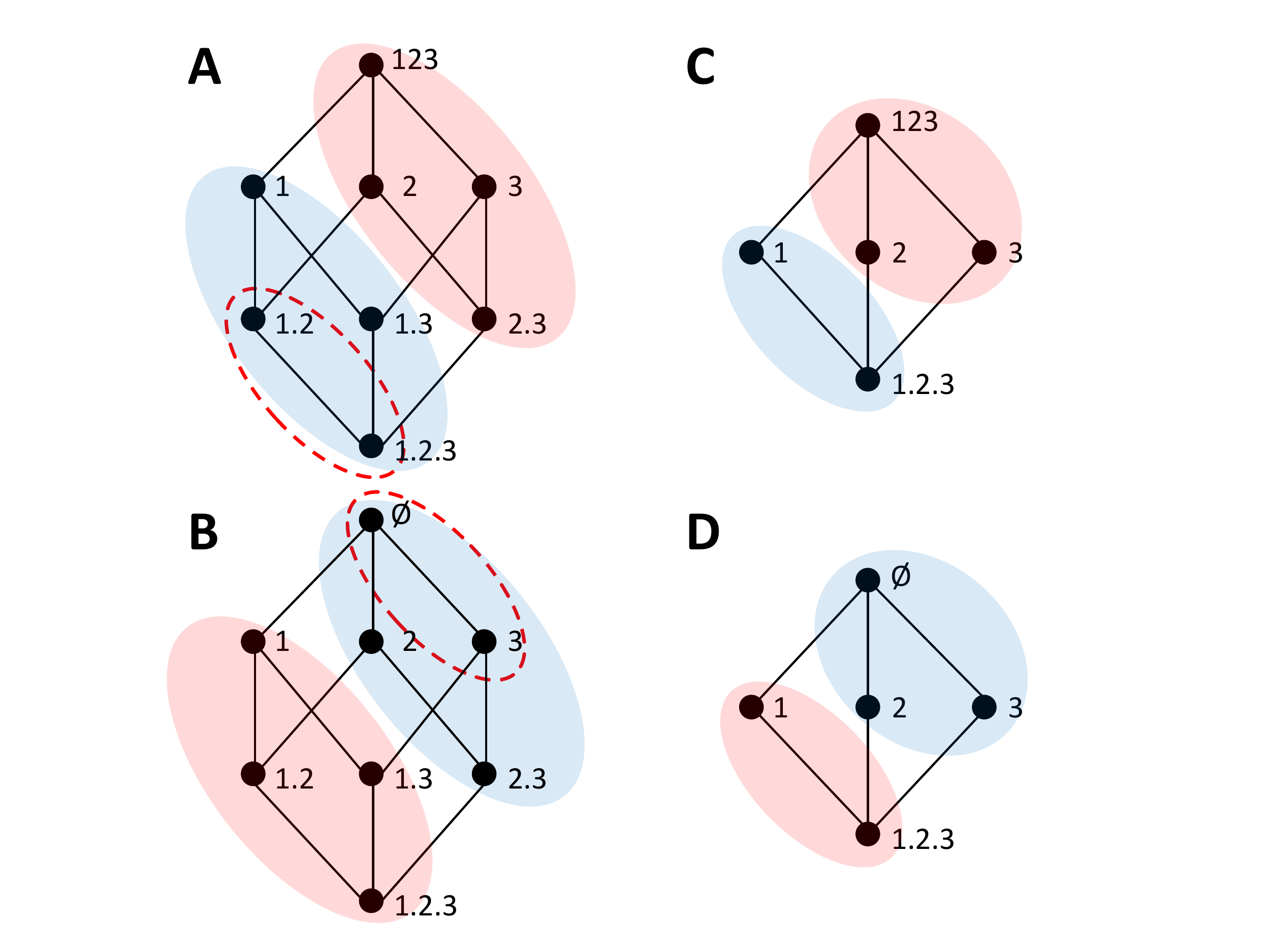}}
  \end{center}
  \caption{Correspondence between information gain and information loss lattices. \textbf{A}, \textbf{C}) Examples of information gain lattices, and their paired information loss lattices (\textbf{B}, \textbf{D}, respectively). The blue shaded areas comprise the collections corresponding to incremental terms that contribute to $I(S;1)$ in each lattice. The pink shaded areas comprise the collections corresponding to incremental terms that contribute to the complementary information $I(S;23|1)$ in each lattice. In \textbf{A}), \textbf{B}), the dashed red lines encircle the incremental terms contributing to $I(S;1.2)$.}
  \label{fig5}
\end{figure}

As another example to gain some intuition about the degree to which gain and loss lattices can be connected, we now reexamine the other two lattices of Figure \ref{fig2}. The blue shaded area of Figure \ref{fig5}A indicates the down set of $1$, containing all the incremental terms accumulated in $I(S;1)$. The complementary set $(\downarrow 1)^C$, indicated by the pink shaded area in Figure \ref{fig5}A, by construction accumulates the remaining information (Eq.\ \ref{s3e5}b), which in this case is $I(S;23|1)$. These two complementary sets of the information gain lattice are mapped to two dual sets in the information loss lattice, as shown in Figure \ref{fig5}B. In Figure \ref{fig5}C we analogously indicate the sets formed by partitioning the gain lattice given the collection $1$, and in Figure \ref{fig5}D the corresponding sets in the information loss lattice. In comparison to the example of Figure \ref{fig4}C,D, for which we already indicated that there is no correspondence between the gain and loss lattices, here in none of the two examples this correspondence is precluded by the difference in the total number of nodes of the gain and loss lattices. However, in Figure \ref{fig5}C,D, the number of nodes is not preserved in the mapping of the partition sets corresponding to collection $1$ from the gain to the loss lattice, which means that the incremental terms cannot be mapped one-to-one from one lattice to the other.

So far we have examined the correspondence of partitions for collections $\alpha$ containing a single source, and hence associated with a directly calculable mutual information, e.g. $I(S;1)$. We have seen that in these cases establishing the correspondence of a partition between lattices is straightforward because the dual sets are identified based on the $\alpha$-partitions of the two lattices, in agreement with Eqs.\ \ref{s3e5}. Accordingly, for these cases in which $\alpha = \mathbf{A}$ is a single source, we can extend Eqs.\ \ref{s3e5}b${, }$d to:

\begin{subequations}
\begin{align}
L(S; \mathbf{A})= I(S;\{ \mathbf{R}\})-I(S; \mathbf{A}) &= \sum_{\beta \in (\downarrow \mathbf{A})^C} \Delta I(S;\beta)\\
I(S; \mathbf{A}')=I(S;\{ \mathbf{R}\})-L(S; \mathbf{A}') &= \sum_{\beta \in (\downarrow \mathbf{A}')^C} \Delta L(S;\beta).
\end{align}
\label{s4e1}
\end{subequations}
However, this direct mapping between the two types of lattices does not hold for collections composed by more than one source. For example, consider the mapping of the cumulative term $I(S;1.2)$, composed by the incremental terms indicated by the dashed red ellipse in Figure \ref{fig5}A. Now in the information loss lattice we cannot take the collection $1.2$ to find the corresponding partition, because the role of the collection $1.2$ in the gain and in the loss lattice is different. $1.2$ indicates the redundant information gain with sources $1$, $2$, and the loss of ignoring other sources apart from $1$, $2$, respectively. To identify the appropriate partition in the information loss lattice we argue that the redundant information between $1$ and $2$ cannot be contained in the accumulated loss of preserving only $1$ or only $2$. Accordingly, $I(S;1.2)$ corresponds to the sum of the incremental terms outside the union of the down sets of $1$ and $2$ in the loss lattice. In general:

\begin{subequations}
\begin{align}
I(S;\alpha) &= \sum_{\beta \in (\bigcup \limits_{\mathbf{B} \in \alpha} \downarrow \mathbf{B})^C} \Delta L(S; \beta),\\
L(S;\alpha') &= \sum_{\beta \in (\bigcup \limits_{\mathbf{B} \in \alpha'} \downarrow \mathbf{B})^C} \Delta I(S; \beta),
\end{align}
\label{s4e2}
\end{subequations}
where the same argument led to relate $L(S;\alpha')$ to gain incremental terms. These equations reduce to Eqs.\ \ref{s4e1} for collections with a single source. It is clear that to connect the cumulative term of a collection $\alpha$ in a type of lattice with a sum of incremental terms in a paired lattice of the other type, the sources composing $\alpha$ must be present as collections in this other lattice. This constrains the lattices that can be paired. However, there is no constraint in the number of incremental terms that are summed to obtain a cumulative term. Therefore, as we actually have illustrated with the examples of Figure \ref{fig4}C,D and \ref{fig5}C,D, for a certain $\alpha$, the number of incremental terms in the sum of Eq.\ \ref{s4e2}a can differ from the number of terms in the sum of Eq.\ \ref{i3}. Similarly, the number of incremental terms can differ between the sums of Eq.\ \ref{s4e2}b and Eq.\ \ref{s3e3}. Plugging Eqs.\ \ref{s4e2}a${, }$b in Eqs.\ \ref{i4}b and \ref{s3e4}b, respectively, we obtain equations relating the incremental terms of the two lattices:

\begin{subequations}
\begin{align}
\Delta I(S; \alpha) &= \sum_{\beta \in (\bigcup \limits_{\mathbf{B} \in \alpha} \downarrow \mathbf{B})^C} \Delta L(S; \beta) - \sum_{k=1}^{|\alpha^-|} (-1)^{k-1} \sum_{
\begin{array}{c} \mathcal{B} \subseteq \alpha^- \\ |\mathcal{B}|=k \end{array}}  \sum_{\beta \in (\bigcup \limits_{\mathbf{B} \in \bigwedge \mathcal{B}} \downarrow \mathbf{B})^C} \Delta L(S; \beta)\\
\Delta L(S; \alpha') &= \sum_{\beta \in (\bigcup \limits_{\mathbf{B} \in \alpha'} \downarrow \mathbf{B})^C} \Delta I(S; \beta) - \sum_{k=1}^{|\alpha'^-|} (-1)^{k-1} \sum_{
\begin{array}{c} \mathcal{B} \subseteq \alpha'^- \\ |\mathcal{B}|=k \end{array}}  \sum_{\beta \in (\bigcup \limits_{\mathbf{B} \in \bigwedge \mathcal{B}} \downarrow \mathbf{B})^C} \Delta I(S; \beta).
\end{align}
\label{s4e3}
\end{subequations}
If the lattices paired are dual, the right hand side of Eq.\ \ref{s4e3}a has to simplify to a single incremental term $\Delta L(S; \beta)$, and similarly the right hand side of Eq.\ \ref{s4e3}b has to simplify to a single incremental term $\Delta I(S; \beta)$. We define duality between information gain and loss lattices imposing this one-to-one mapping of the incremental terms:

\paragraph{\textbf{Lattice duality}:} An information gain lattice associated with a set $\mathcal{C}$ and an information loss lattice associated with a set $\mathcal{C}'$, built according to the ordering relations of Eqs.\ \ref{i2}, \ref{s3e2}, and fulfilling the constraints of Eqs.\ \ref{i3}, \ref{i4}, \ref{s3e3}, \ref{s3e4}, are dual if and only if

\begin{subequations}
\begin{align}
& \forall \alpha \in \mathcal{C}\  \exists \beta \in \mathcal{C}': \Delta I(S; \alpha) = \Delta L(S; \beta),\\
& \forall \alpha \in \mathcal{C}\  \exists \beta \in \mathcal{C}': I(S; \alpha) = \sum_{\gamma \in \downarrow \alpha} \Delta I (S;\gamma) = \sum_{\gamma \in \uparrow \beta} \Delta L(S;\gamma),\\
& \forall \alpha' \in \mathcal{C}'\  \exists \beta' \in \mathcal{C}: \Delta L(S; \alpha') = \Delta I(S; \beta'),\\
& \forall \alpha' \in \mathcal{C}'\  \exists \beta' \in \mathcal{C}: L(S; \alpha') = \sum_{\gamma \in \downarrow \alpha'} \Delta L (S;\gamma) = \sum_{\gamma \in \uparrow \beta'} \Delta I(S;\gamma).
\end{align}
\label{s4e4}
\end{subequations}

\begin{figure}
  \begin{center}
    \scalebox{0.55}{\includegraphics*{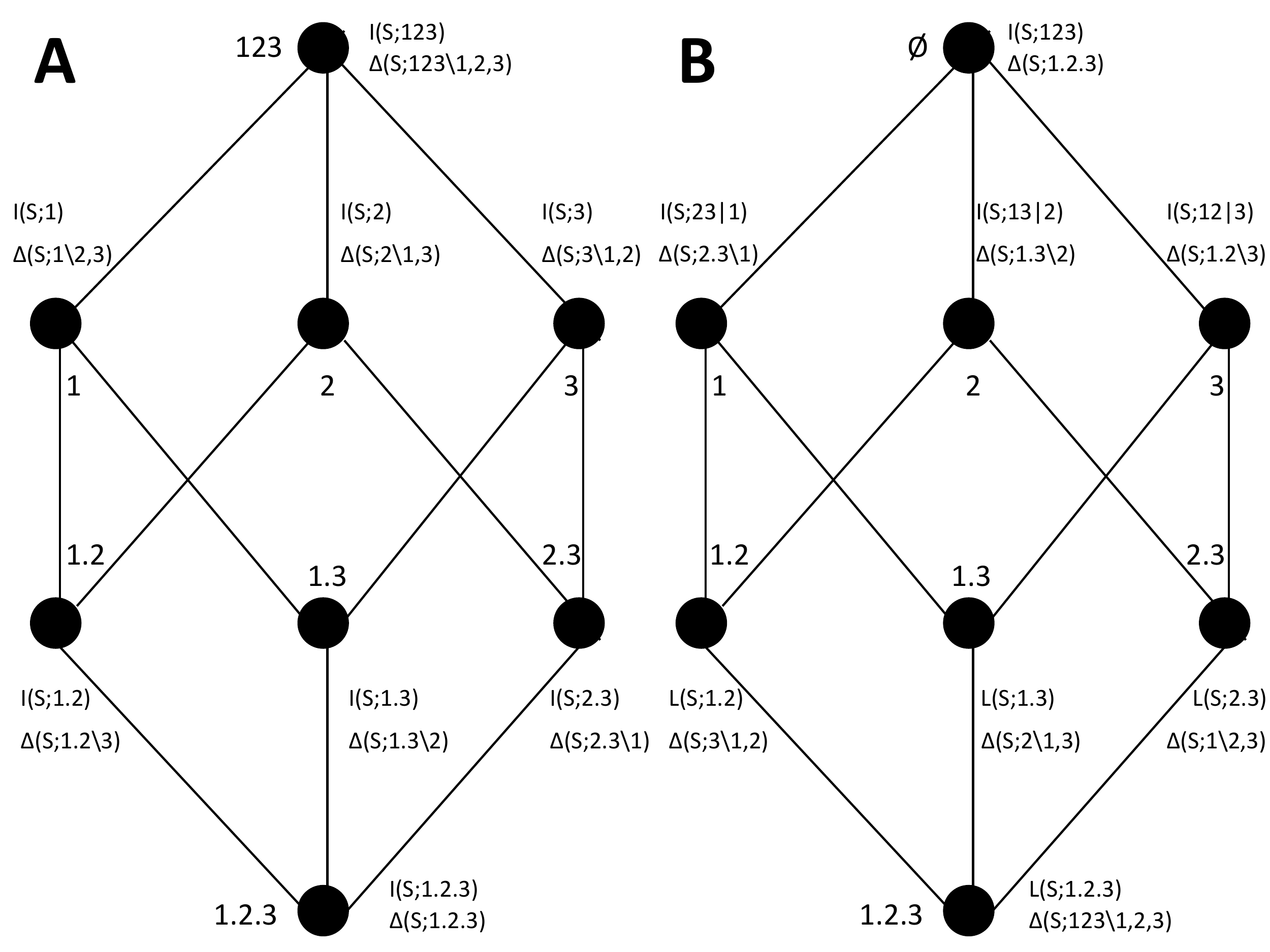}}
  \end{center}
  \caption{Dual trivariate decompositions for the sets of collections that do not contain bivariate sources. \textbf{A}) Information gain lattice.  \textbf{B}) Information loss lattice. In each node together with the collection the corresponding cumulative and incremental terms are indicated. Note that the incremental terms are common to both lattices and can be mapped by reversing the lattice up/down and right/left. In the information loss lattice the cumulative terms of collections containing single sources, $L(S;i),\ i=1, 2, 3$, are directly expressed as the corresponding conditional information.}
  \label{fig6}
\end{figure}

This definition does not provide a procedure to construct the dual information loss lattice from an information gain lattice, or viceversa. However, we have found and we here conjecture that a necessary condition for two lattices to be dual is that they contain the same collections except $\{\mathbf{R}\}$ at the top of the gain lattice being replaced by $\emptyset$ at the loss lattice. In particular, the lattices constructed from the full domain of collections, $\mathcal{A} \{ \mathbf{R} \}$ for the gain and $\mathcal{A}^* \{ \mathbf{R} \}$ for the loss, are dual. In Figure \ref{fig6} we show an example of dual lattices, the pair already discussed in Figure \ref{fig5}A,B. We detail all the cumulative and incremental terms in these lattices. While the cumulative terms are specific to each lattice, the incremental terms, in agreement with Eqs.\ \ref{s4e4}a$, $c, are common to both. In more detail, the incremental terms are mapped from one lattice to the other by an up/down and right/left reversal of the lattice. From these two reversals, the right/left is purely circumstantial, a consequence of our choice to locate the collections common to both lattices in the same location (for example, to have the collections ordered $1$, $2$, $3$ in both lattices instead of $3$, $2$, $1$ for one of them). Oppositely, the up/down reversal is inherent to the duality between the lattices and reflects the relation between the summation in down sets or up sets in the summands of Eqs.\ \ref{s4e4}b$, $d.

\begin{table}
\centering
\begin{tabular}{| l c | c |}
\hline \hline
Term &  Measure\\
\hline \hline
$\Delta (S; i.j.k) = I(S; i.j.k)$ & $ \begin{array}{c} \min \limits_{i.j.k} I(S;ijk) - \min \limits_{i.j} I(S;j|i) \\  - \min \limits_{i.k} I(S;i|k) -\min \limits_{j.k} I(S;k|j) \end{array} $ \\
\hline
$\Delta (S; i.j \backslash k)$ & $ \begin{array}{c} \min \limits_{i.k} I(S;ijk) + \min \limits_{j.k} I(S;ijk) \\  - \min \limits_{i.j.k} I(S;ijk) -I(S;k) \end{array} $\\
\hline
$\Delta (S; i \backslash j, k )$ & $\min \limits_{i.j.k} I(S;ijk) - \min \limits_{j.k} I(S;ijk)$ \\
\hline
$\Delta (S; i j k \backslash i, j, k ) = L(S;i.j.k)$ & $I(S;ijk)- \min \limits_{i.j.k} I(S;ijk)$\\
\hline
$I(S;i.j)$ & $I(S; j)- \min \limits_{i.j} I(S; j|i)$ \\
\hline
$L(S;i.j)$ & $I(S;ijk) - \min \limits_{i.j} I(S;ijk)$\\
\hline
$L(S;i)$  & $I(S;j k|i)$\\
\hline
\end{tabular}
\label{table1}
\caption{Components of the mutual information dual decompositions of Figure \ref{fig6} based on the synergy measure defined in \cite{Bertschinger12}.}
\end{table}

To provide a concrete example of information gain and information loss dual decompositions we here adopted and extended to the multivariate case the bivariate synergy measure defined in \cite{Bertschinger12}. Table $1$ lists all the resulting expressions when this measure is used to determine all the terms in both decompositions. The measure associated with terms $L(S;i.j)$ corresponds to the original bivariate measure of synergy of \cite{Bertschinger12}. This measure is extended in a straightforward way to the multivariate case, and in particular for the trivariate case corresponds to the term $L(S;i.j.k)$. The bivariate redundancy measure also already used in \cite{Bertschinger12} corresponds to $I(S;i.j)$. The rest of incremental terms can be obtained from the information loss lattice using Eq.\ \ref{s3e4}. Note that we could have proceeded in a similar way starting from a definition of the cumulative terms in the gain lattice, such as $I_{min}$, and then determining the terms of the loss lattice. Here we use this concrete decomposition only as an example and it is out of the scope of this work to characterize the properties of the resulting terms. Alternatively, we focus on discussing the properties related with the duality of the decompositions.

Most importantly, the dual lattices provide a self-consistent quantification of synergy and redundancy. Eqs.\ \ref{s4e4}a$, $c, together with the fact that the bottom incremental terms of lattices are also cumulative terms, ensure that, combining different dual lattices of different order $n$ and composed by different subsets, as studied in Section \ref{s2}, all incremental terms correspond to a bottom cumulative term of a certain lattice. For example, for the lattices of Figure \ref{fig6}, the bottom cumulative term in the information gain lattice, the redundancy $I(S;i.j.k)$, is equal to the top incremental term of the loss lattice, $\Delta (S; i.j.k)$. Similarly, the bottom cumulative term of the loss lattice, the synergy $L(S; i.j.k)$, is equal to the top incremental term of the gain lattice $\Delta (S; ijk \backslash i,j,k)$.

For dual lattices, the iterative procedure of Section \ref{s2_2} can be applied to recover the components of the information gain lattice from a definition of synergy and the components calculated in this way are equal to the ones obtained from the mapping of incremental terms from one lattice to the other. In more detail, let us refer to the bottom and top terms by $\bot$ and $\top$, respectively, and distinguish between generic terms such as $I(S; \alpha)$ and a specific measure assigned to it, $\bar{I}(S; \alpha)$. One can define the synergistic top incremental term of the gain lattice using the measure assigned to the bottom cumulative term of the loss lattice, imposing $\Delta I (S; \top) \equiv \bar{L}(S; \bot)$ and self-consistency ensures that the measures obtained fulfill $\bar{I}(S; \bot) = \Delta \bar{L}(S; \top)$. Similarly, self-consistency assures that, if one takes as a definition of redundancy for the cumulative terms of the gain lattice the measure assigned to the incremental terms of the loss lattice based on a definition of synergy, consistent incremental terms are obtained in the gain lattice. That is, $I(S; \bot) \equiv \Delta \bar{L}(S; \top)$ results in $\Delta \bar{I} (S; \top) = \bar{L}(S; \bot)$. It can be checked that these self-consistency properties do not hold in general, for example for the lattices of Figure \ref{fig5}C,D. The properties of dual lattices guarantee that, within the class of dual lattices connected by the decomposition-invariance of cumulative terms, inconsistencies of the type discussed in Section \ref{s2_2} do not occur, and all the terms in the decompositions are not decomposition-dependent.

\section{Discussion}

In this work we extended the framework of \cite{Williams10} focusing on the lattices that underpin the mutual information decompositions. We started generalizing the type of information gain lattices introduced by \cite{Williams10}. By considering more generally which information gain lattices can be constructed (Section \ref{s2_1}), we reexamined the constraints that \cite{Williams10} identified for the lattice's components (Eq.\ \ref{i1}) and ordering relation (Eq.\ \ref{i2}). These constraints were motivated by the link of each node in the lattice with a measure of accumulated information. We argued that it is necessary to check the validity of each specific lattice given each specific set of variables. We indicated that this checking can overcome the problems found by \cite{Rauh14} with the original lattices described in \cite{Williams10}. In particular, we showed that the existence of nonnegative components in the presence of deterministic relations between the variables is directly a consequence of the non-compliance of the validity constraints.

For our generalized set of information gain lattices, we examined the relations between the terms in different lattices (Section \ref{s2_1}). We pointed out that the two types of information-theoretic quantities associated with the lattices have different invariance properties: The cumulative terms of the information gain lattices are invariant across decompositions, while the incremental terms are decomposition-dependent and are only connected across lattices through the relations resulting from the invariance of the cumulative terms. This produces a qualitative difference in the properties of the redundancy components of the decompositions, which are associated with cumulative terms in the information gain lattices, and the unique or synergy components, which correspond to incremental terms. This difference has practical consequences when trying to construct a mutual information decomposition from a measure of redundancy or a measure of synergy or unique information, respectively. In the former case, as described in \cite{Williams10}, the terms of the decomposition can be derived straightforwardly given that the redundancy measure identifies the cumulative terms. In the latter, for the multivariate case, it is not straightforward to construct the decomposition because the synergy or uniqueness measures only allow identifying specific incremental terms. Exploiting the connection between different lattices that results from the invariance of the cumulative terms, we proposed a procedure to generally construct information gain decompositions from a measure of synergy or unique information (Section \ref{s2_2}). This procedure allows applying to the multivariate case measures of synergy \citep{Griffith13} or unique information \citep{Bertschinger12} for which associated decompositions had only been constructed for the bivariate case. However, the application of this procedure led us to recognize inconsistencies in the determination of decompositions components across lattices. We argued that these inconsistencies are a consequence of the intrinsic decomposition-dependence of synergy and unique information components, inherited from their correspondence to incremental terms in the information gain lattice.

We then introduced an alternative decomposition of the mutual information based on information loss lattices (Section \ref{s3}). The role of redundancy and synergy components is exchanged in the loss lattices with respect to the gain lattices, now being the synergy components the ones associated with the cumulative terms. We defined the information loss lattices analogously to the gain lattices, determining validity constraints for the components and introducing an ordering relation to construct the lattices. Cumulative and incremental terms are related in the same way as in the gain lattices, establishing the connection between the lattice and the mutual information decomposition. This type of lattices allows readily determining the information decomposition from a definition of synergy. Furthermore, the information loss lattices can be useful in relation to other alternative information decompositions \citep{Schneidman03b, Olbrich15, Perrone16}. However, analogous inconsistencies to the ones found for the information gain lattices affect now the redundancy components, which now correspond to incremental terms. Therefore, we studied in general the correspondence between information gain and information loss lattices, in order to determine how to jointly quantify synergy and redundancy. The final contribution of this work was the definition of dual gain and loss lattices (Section \ref{s4}). Within a dual pair, the gain and loss lattices share the incremental terms, which can be mapped one-to-one from the nodes of one lattice to the other. Duality ensures self-consistency, so that the redundancy components obtained from a synergy definition are the same as the synergy components obtained from the corresponding redundancy definition.

As in the original work of \cite{Williams10} that we aimed to extend, we have here considered generic variables, without making any assumption about their nature and relations. A case which however deserves special attention is that of variables associated with time-series, so that information decompositions allow studying the dynamic dependencies in the system \citep{Chicharro12b, Faes15}. Practical examples include the study of multiple-site recordings of the time course of neural activity at different brain locations, with the aim of understanding how information is processed across neural systems \citep{Valdes11}. In such cases of time-series variables, a widely-used type of mutual information decomposition aims to separate the contribution to the information of different causal interactions between the subsystems \citep[e.\ g.\ ][]{Solo08, Chicharro11b}. Considering synergistic effects is also important when trying to characterize the causal relations \citep{Stramaglia14}. In fact, when causality is analyzed by quantifying statistical predictability using conditional mutual information, a link between these other decompositions and the one of \cite{Williams10} can be readily established \citep{Williams11, Lizier13}.

The proposal of \cite{Williams10} has proven to be a fruitful conceptual framework and connections to other approaches to study information in multivariate systems have been explored \citep{Wibral15, Banerjee15, James16}. However, despite subsequent attempts \citep[e.\ g.\ ][]{Harder12, Bertschinger12, Griffith13, Ince16}, it is still an open question how to decompose in multivariate systems the mutual information into nonnegative contributions that can be interpreted as synergy, redundancy, or unique components. This issue constitutes the main challenge that limits so far the practical applicability of the framework. Other challenges for this type of decompositions are to be able to further relate the terms in the decomposition with a functional description of the parts composing the system \citep{Panzeri17} and, in the case of dynamic systems, to adapt the decompositions to incorporate an interventional instead of only statistical predictability approach to causality \citep{Chicharro14, Panzeri17}. This situation is, in practice, relevant for example to dissect information transmission in neural circuits during behavior, which can be done combining the analysis of time-series recordings of neural activity using information decompositions with space-time resolved interventional approaches based on brain perturbation techniques such as optogenetics \citep{Oconnor13, Otchy15, Panzeri17}. This interventional approach can be incorporated to the framework by adopting interventional information-theoretic measures suited to quantify causal effects \citep{Ay06, Lizier10, Chicharro12}. The work that we have presented here does not address yet these challenges. However, overall, this work provides a wider perspective to the ground constituents of the mutual information decompositions introduced by \cite{Williams10}, introduces new types of lattices, and helps to clarify the relation between synergy and redundancy measures with the lattices components. The consolidation of this theoretical framework is expected to foster future applications.

\paragraph{\textbf{Acknowledgments}}

This work was supported by the Fondation Bertarelli and by the Autonomous Province of Trento, Call 'Grandi Progetti $2012$', project 'Characterizing and improving brain mechanisms of attention-ATTEND'. We are grateful to E. Piasini, H. Safaai, G. Pica, J. Bim, R. Ince, and V. De Feo for useful discussions on these topics.

\paragraph{\textbf{Author’s Contribution}}

All authors contributed to the design of the research. The research was carried out by Chicharro. The manuscript was written by Chicharro and Panzeri. All authors read and approved the final manuscript.

\paragraph{\textbf{Conflicts of Interest}}

The authors declare no conflict of interest.

\vspace*{7mm}

\appendix{}

\section{Lattice theory definitions}
\label{a1}

We here review some concepts of lattice theory and of the construction of information decompositions based on collections lattices. For further review and references to specialized textbooks see \cite{Williams10}.

\paragraph{\textbf{Definition 1}:} A pair $\langle X, \leq \rangle$ is a partially ordered set or poset
if $\leq$ is a binary relation on $X$ that is reflexive, transitive and antisymmetric.

\paragraph{\textbf{Definition 2}:} Let $\langle X, \leq \rangle$ be a poset, and let $Y \subseteq X$. An
element $x \in X$ is a lower bound for $Y$ if $\forall y \in Y, y \geq x$. An upper bound for Y is defined dually.

\paragraph{\textbf{Definition 3}:} An element $x \in X$ is the greatest lower bound or
infimum for $Y$ , denoted $\mathrm{inf}\ Y$, if $x$ is a lower bound of $Y$ and $ \forall y \in Y$ and $\forall z \in X; y \geq z$ implies $x \geq z$. The least upper bound  or supremum for $Y$ , denoted $\mathrm{sup}\ Y$, is defined dually.

\paragraph{\textbf{Definition 4}:} A poset $\langle X, \leq \rangle$ is a lattice if, and only if, $\forall x, y \in X$ both $\mathrm{inf}\ \{ x, y\}$ and $\mathrm{sup}\ \{ x, y\}$ exist in $X$. For $Y \subseteq X$, we use $\bigwedge Y$ and $\bigvee Y$ to denote the infimum and supremum of all elements in $Y$, respectively.

\paragraph{\textbf{Definition 5}:} For $a, b \in X$, we say that $a$ is covered by $b$ if $a < b$ and $a \leq c < b \Rightarrow a = c$. The set of elements that are covered by $b$ is denoted by $b^-$.

\paragraph{\textbf{Definition 6}:} For any $x \in X$, the down-set of $x$ is the set $ \downarrow x = \{y \in X : y \leq x \} $. The up-set $\uparrow x$ of $x$ is defined analogously.

Apart from these definitions from lattice theory we here introduce, as a concept more specific of the information decompositions, the concept of increment sublattice:

\paragraph{\textbf{Definition 7}:} For a lattice built with the collections set $\mathcal{C}$, for any $\alpha \in \mathcal{C}$, the increment sublattice is $\diamondsuit \alpha = \{ \bigwedge \mathcal{B}: \mathcal{B} \subseteq \alpha^-, |\mathcal{B}|=k, k = 1,..., |\alpha^-|\}$ .

\section{Validity checking to overcome the nonnegativity counterexample of \cite{Rauh14}}
\label{a2}

We here examine in more detail the nonnegativity counterexample studied in \cite{Rauh14} that we mentioned in Section \ref{s2_1}. In this example two variables $Y_1 Y_2$ are independently uniformly distributed binary variables, and a third is generated as $Y_3 = Y_1\ XOR\ Y_2$. Furthermore, $\mathbf{S} = (Y_1, Y_2, Y_3)$. The variables have deterministic relations, such that any pair $\{ Y_i, Y_j\}, i \neq j$ determines the third. We start by reviewing their arguments. The identity axiom proposed by \cite{Harder12} imposes that $I(Y_i Y_j; Y_i.Y_j)= I(Y_i;Y_j)=0\ \mathrm{bit}, i \neq j$. Given the deterministic relation between the variables this implies that $I(\mathbf{S}; Y_i.Y_j)=0\ \mathrm{bit}, i \neq j$. By monotonicity ascending the lattice of Figure \ref{fig1}B, also $I(\mathbf{S}; Y_1.Y_2.Y_3)=0$ bit. Accordingly, also the incremental terms of the corresponding nodes vanish. In the next level of the gain lattice, $I(\mathbf{S}; Y_i.Y_j Y_k)= I(Y_1 Y_2 Y_3; Y_i.Y_j Y_k)$ and hence applying again the identity axiom, $I(\mathbf{S}; Y_i.Y_j Y_k)= I(Y_i;Y_j Y_k)=1$ bit. This also leads to $\Delta I(\mathbf{S}; Y_i.Y_j Y_k \backslash Y_j, Y_k)=1$ bit. Furthermore, by monotonicity, $I(\mathbf{S}; Y_1 Y_2.Y_1 Y_3.Y_2 Y_3) \leq I(\mathbf{S}; Y_1 Y_2 Y_3)=2$ bit. This leads to the incremental term $\Delta I(\mathbf{S}; Y_1 Y_2.Y_1 Y_3.Y_2 Y_3 \backslash Y_1, Y_2, Y_3) \leq 2-3\ \mathrm{bit} = -1 \mathrm{bit}$. Since this derivation is based on the axioms and not on the specific properties of the measures used, this proves that, for the lattice of Figure \ref{fig1}B and for this specific set of variables, there is no measure that can be used to define the terms in the decomposition so that nonnegativity is respected.

We completely agree with the derivation of \cite{Rauh14}. What we argue is that in this case the non-compliance of nonnegativity is a direct consequence of how the deterministic relations between the variables render some of the collections that form part of the lattice of Figure \ref{fig1}B invalid according to the constraints that define the domain of collections (Eq.\ \ref{i1}), and render some ordering relations invalid according to the ordering rule of Eq.\ \ref{i2}. Therefore, adopting the generalized framework that we have proposed, this counterexample can be reinterpreted by saying that the full lattice is not valid for these variables, but that still other lattices are possible. In particular, for the lattice of Figure \ref{fig1}B, one can use the deterministic relations between the variables to substitute each bivariate source $Y_i Y_j$ by $Y_1 Y_2 Y_3$, and then check which collections are invalid. After removing these invalid collections and rebuilding the edges between the remaining collections according to the ordering relation, the lattice of Figure \ref{fig1}D is obtained.

However, it can be checked, following a derivation analogous to the one of \cite{Rauh14}, that also for the lattice of Figure \ref{fig1}D nonnegativity is not accomplished, in particular by $I(\mathbf{S}; Y_1 Y_2 Y_3\backslash Y_1, Y_2, Y_3)$. This is because, still by the deterministic relations, the top collection could be reduced to any collection $Y_i Y_j$. In contrast to the lattice of Figure \ref{fig1}B, in Figure \ref{fig1}D this reduction would not led to a duplication of a collection, since no bivariate sources are present in other nodes, but it still invalidates the ordering relations in the lattice. In particular, if $Y_1 Y_2 Y_3$ is replaced by $Y_i Y_j$, the edge between $Y_i Y_j$ and $Y_k$ has to be removed. The remaining structure is not a lattice anymore, given the Definition $4$ in Appendix $A$. In Appendix $C$ we briefly discuss more general information decompositions for structures that are not lattices, but here we still restrict ourselves to lattices. Within the set of lattices, it is now clear that in this case the deterministic relations render invalid any lattice containing the three variables, and thus only lattices analogous to the one of Figure \ref{fig1}A can be built. For these lattices with two variables, $I(\mathbf{S}; Y_i.Y_j)=0$ bit, $I(\mathbf{S}; Y_i)=1$ bit, and $I(\mathbf{S}; Y_i Y_j)= 2$ bit lead to all incremental terms being nonnegative. Instead of a counterexample of the nonnegativity of the incremental terms, we can interpret this case as an example in which the relations between the variables invalidate certain lattices. The possibility to generally construct multivariate nonnegative decompositions, even after these validity checking, remains an open question.

\section{The requirements for the nonnegativity of the decomposition incremental terms}
\label{a3}

We here review the proofs of Theorems $3-5$ of \cite{Williams10} from a general perspective, identifying their key ingredients. The aim is to recognize which constraints exist to further generalize the type of structures that can be used to build mutual information decompositions while preserving the same relation between the structures and the information-theoretic terms. Furthermore, we want to identify the properties required to ensure nonnegativity for the incremental terms, and assess the degree to which these properties can be shared by other measures or are mainly specific of the form of the measure $I_{min}$ proposed in \cite{Williams10}. This is important because the proposal of \cite{Williams10} is the only one in which nonnegativity of the decomposition components has been proven for the multivariate case. This appendix does not aim to be fully autonomous and assumes the previous reading of the proofs in \citep{Williams10}.

We start discussing Theorem $3$ of \cite{Williams10}. The theorem states the expression for the incremental terms of the information gain lattices that we indicated in Eq.\ \ref{i4}. The expression of Eq.\ \ref{i4}a results directly from the implicit definition of the incremental terms in Eq.\ \ref{i3} and does not require that the structure formed by the collections given the ordering relation is a lattice. Conversely, Eq.\ \ref{i4}b requires that, at least for the elements in $\diamondsuit \alpha$, the structure forms a lattice, namely the increment sublattice. Although \cite{Williams10} formulated this theorem specifically for $I_{min}$, it does not depend on the properties of the measure and relies only on the lattice properties and the connection between the lattice and the information decomposition given by Eq.\ \ref{i3}. This is why we can use the expressions of Eq.\ \ref{i4} without any specification about the form of the mutual information measures used to build the decomposition.

We now consider Theorem $4$ of \cite{Williams10}. For the proof of this theorem not only lattice properties but also the properties of $I_{min}$ were used. We are interested in separating which of these properties correspond to the axioms generically required for any measure of redundancy \citep[e.\ g.\ ][]{Griffith14}, and which are specific of the form of $I_{min}$. First, the proof uses Theorem $3$ and Lemma $2$ of \cite{Williams10}, which do not depend on the specific properties of $I_{min}$, nor in any generic axiom for redundancy measures. Note however that the proof uses Eq.\ \ref{i4}b and not only Eq.\ \ref{i4}a to express the incremental terms as a function of cumulative terms, and thus, for a certain $\alpha$, only holds if the structure is compatible with a lattice for $\diamondsuit \alpha$. Second, the proof relies on a very specific property of the form of $I_{min}$: For a given collection, this measure is defined based on a minimum operation acting on a set of values, each value associated with one of the sources contained in the collection. In more detail, each value corresponds to the Specific Information for the corresponding source, and thus it is nonnegative and monotonicity holds between sources with more variables. This means that, when considering each summand in $I_{min}$ for $S=s$, a cumulative term $I(S=s; \alpha)$ is a function of the cumulative terms associated with the collections formed by each of the sources in $\alpha$ alone. This is relevant because it allows relating the measures in each node of the lattice beyond the generic relations characteristic of the decomposition. In more detail, in the proof it allows substituting a minimum operation acting on the sources contained in the infimum of a set of collections by two minimum operations, acting on the collections in that set and on the sources in each of these collections, respectively.

Finally, Theorem $5$, which proofs the nonnegativity of the incremental terms, relies on Theorem $4$, the nonnegativity of cumulative terms $I(S; \alpha)$, and monotonicity of the Specific Information. Overall, we see that the specific closed form expression of the incremental terms stated in Theorem $4$ is fundamental to prove the nonnegativity of the incremental terms. The key property of $I_{min}$ to prove Theorem $4$ does not follow from the generic axioms proposed for redundancy measures, and is not shared by other measures that have been proposed \citep[e.\ g.\ ][]{Harder12, Griffith13, Bertschinger12}. This renders the proof of Theorem $4$ and $5$ specific to $I_{min}$, in contrast to the proof of Theorem $3$. Accordingly, our reexamination of the proofs of \cite{Williams10} helps to point out that any attempt to prove the nonnegativity of the mutual information decomposition based on an alternative measure cannot in general follow the same procedure.

\begin{figure}
  \begin{center}
    \scalebox{0.55}{\includegraphics*{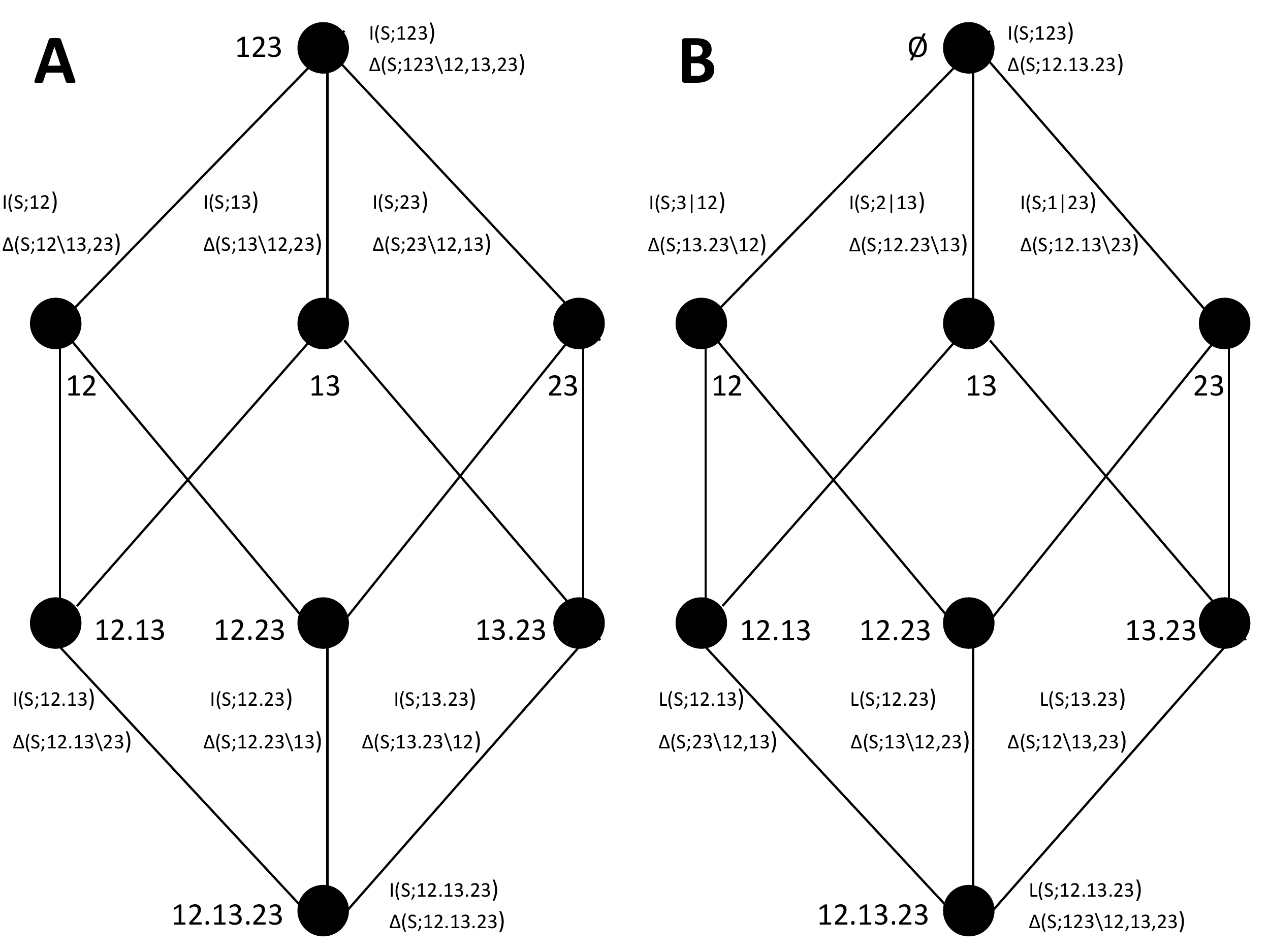}}
  \end{center}
  \caption{Analogous to Figure \ref{fig6} but for the trivariate decomposition based only on collections that do not contain univariate sources.}
  \label{fig7}
\end{figure}

\section{Another example of dual decompositions}
\label{a4}

As a second example of a pair of dual decompositions we show in Figure \ref{fig7}, also for the case of three variables, the decompositions for the sets of collections that do not contain univariate sources.


\end{document}